\documentclass[12pt,preprint]{aastex}

\usepackage{amsmath}

\def\p/{\mbox{$^1$}}
\def\pp/{\mbox{$^2$}}
\def\ppp/{\mbox{$^3$}}
\def\pppp/{\mbox{$^4$}}
\def\m/{\mbox{$^{-1}$}}
\def\mm/{\mbox{$^{-2}$}}
\def\mmm/{\mbox{$^{-3}$}}
\def\mmmm/{\mbox{$^{-4}$}}
\def\Ms/{\mbox{M$_\odot$}}

\def\ebv{\mbox{$E(4405-5495)$}}
\def\rv{\mbox{$R_{5495}$}}

\def\xo{\mbox{$x_{\rm o}$}}
\def\fx{\mbox{$f_X(x)$}}
\def\fxo{\mbox{$f_{X_{\rm o}}(\xo)$}}
\def\fxocx{\mbox{$f_{X_{\rm o}|X}(\xo|x)$}}
\def\fxox{\mbox{$f_{X_{\rm o},X}(\xo,x)$}}
\def\moLLL{\mbox{$m_{\rm o,LLL}$}}
\def\neff{\mbox{$n_{\rm eff}$}}
\def\mstav{\mbox{$m_{\ast,\rm av}$}}
\def\mo{\mbox{$m_{\rm o}$}}
\def\mr{\mbox{$m_{\rm r}$}}
\def\ml{\mbox{$m_{\rm lower}$}}
\def\mu{\mbox{$m_{\rm upper}$}}
\def\sr{\mbox{$\sigma_{\rm r}$}}
\def\fm{\mbox{$f_M(m)$}}
\def\ns{\mbox{$n_{\rm s}$}}
\def\fmo{\mbox{$f_{M_{\rm o}}(\mo)$}}
\def\fmocm{\mbox{$f_{M_{\rm o}|M}(\mo|m)$}}
\def\fmom{\mbox{$f_{M_{\rm o},M}(\mo,m)$}}
\def\sk{\mbox{$\gamma_1$}}
\def\skg{\mbox{$\gamma_{1,\Gamma}$}}
\def\ku{\mbox{$\gamma_2$}}
\def\kug{\mbox{$\gamma_{2,\Gamma}$}}
\def\gammap{\mbox{$\gamma^\prime$}}
\def\dgamma{\mbox{$\Delta\gamma$}}
\newcommand{\sci}[2]{\mbox{$#1\cdot 10^{#2}$}}

\shorttitle{CMF biases created by uncertainties in the observed mass}
\shortauthors{Ma\'{\i}z Apell\'aniz}

\slugcomment{Accepted for publication in the Astrophysical Journal}

\begin{document}

\title{Biases on initial mass function determinations. III. Cluster masses derived from unresolved photometry}

\author{J. Ma\'{\i}z Apell\'aniz\altaffilmark{1,2}}
\affil{Instituto de Astrof\'{\i}sica de Andaluc\'{\i}a-CSIC, Camino bajo de Hu\'etor 50, 18008 Granada, Spain}


\altaffiltext{1}{e-mail contact: {\tt jmaiz@iaa.es}.}
\altaffiltext{2}{Ram\'on y Cajal fellow.}

\begin{abstract}
It is currently common to use spatially unresolved multi-filter broad-band photometry to determine the masses of individual stellar clusters (and hence the cluster mass function, CMF). I analyze the stochastic 
effects introduced by the sampling of the stellar initial mass function (SIMF) in the derivation of the individual masses and the CMF and I establish that such effects are the largest contributor to the
observational uncertainties. An analytical solution, valid in the limit where uncertainties are small, is provided to establish the range of cluster masses over which the CMF slope can be obtained with a given 
accuracy. The validity of the analytical solution is extended to higher mass uncertainties using Monte Carlo simulations and the Gamma approximation. The value of the Poisson mass is calculated for a
large range of ages and a variety of filters for solar-metallicity clusters measured with single-filter photometry. A method that uses the code
CHORIZOS is presented to simultaneously derive masses, ages, and extinctions. The classical method of using unweighted $UBV$ photometry to simultaneously establish ages and extinctions of stellar clusters is 
found to be unreliable for clusters older than $\approx 30$ Ma, even for relatively large cluster masses. On the other hand, augmenting the filter set to include longer-wavelength filters and using weights for 
each filter increases the range of masses and ages that can be accurately measured with unresolved photometry. Nevertheless, a relatively large range of masses and ages is found to be dominated by SIMF sampling 
effects that render the observed masses useless, even when using $UBVRIJHK$ photometry. A revision of some literature results affected by these effects is presented and possible solutions for future observations
and analyses are suggested.
\end{abstract}


\section{Introduction}

	This paper is the third one of a series where we explore the effects of different biases on the determination of the stellar and cluster 
mass functions (SMFs and CMFs, respectively). In paper I \citep{MaizUbed05} we analyzed the numerical biases induced by 
using bins of equal width when fitting power-laws to binned data (an effect that is more general than its application to the calculation of mass functions).
Those biases can be eliminated in several ways, of which a simple one is by grouping the data in
equal-number bins (as opposed to equal-width bins). In paper II \citep{Maiz08a} I explored the effect of unresolved multiple systems, either physical or
chance alignments, especially for the high-mass end of the stellar initial mass function (SIMF). In this paper I analyze the effect of random uncertainties 
in the mass determinations of individual stellar clusters and on the global properties of the obtained CMF as derived from spatially integrated (i.e. unresolved) 
photometry. I am currently working on the fourth paper of the series, which will explore the same issues as this one but referred to stellar instead of cluster 
masses.

	The measurement of the masses of unresolved stellar clusters in external galaxies has become popular in the last decade, especially thanks to the 
availability of HST imaging (see \citealt{ZhanFall99,Lars02,deGretal05,Ubedetal07b,Doweetal08} for examples). Accurately measuring stellar cluster masses 
is crucial to understand their evolution and, more specifically, their destruction rates and mechanisms. This is usually done by analyzing a large ensemble of
clusters within a galaxy and deriving the present-day CMFs as a function of age. Current results regarding 
how and at what rate stellar clusters are destroyed are inconclusive, with two alternative empirical models being proposed in the literature (\citealt{Lame08} and 
references therein). 

	One outstanding issue with the calculation of stellar cluster masses is SIMF sampling. In a series of papers, Miguel Cervi\~no and his collaborators 
(\citealt{CervLuri06} and references therein) have shown that for clusters with less than a certain number of stars the SIMF is not well sampled and, as a 
consequence, there are added uncertainties in the derivation of cluster properties from unresolved photometric or spectroscopic data. Thus, two clusters of the 
same mass, age, and metallicity can have quite different integrated properties because of the differences in their initial stellar population caused by the 
stochastic nature of star formation\footnote{Note that other factors such as the dynamical evolution due to both internal and external causes can and indeed frequently 
do induce differences between otherwise initially similar clusters. However, those will not be considered in this paper.}. Undoubtedly, it is important to determine 
whether such effects are distorting the analysis of stellar cluster evolution by inducing biases in the observed CMFs.

	The papers by Cervi\~no et al. deal mostly with the theoretical issues of SIMF sampling and provide a framework for its general study. The goal of this paper is 
more limited but at the same time more practical. It is more limited because I will concentrate on one observable, the cluster mass, and will deal  with others (e.g. age) only
as long as they influence the value of the observed mass. Furthermore, for the sake of simplicity I will restrict the analysis to solar-metallicity clusters.
The practical character of this paper arises from its main aims, which are to provide [a] specific criteria to determine when cluster masses 
can be obtained, [b] methods for a more precise measurement of individual cluster masses, and [c] corrections to eliminate biases in both the individual masses and the 
CMF. In a sense, this paper can be understood as an unresolved-population equivalent to the resolved-stellar-population methods and tools used for extracting unbiased ages or
star-formation histories from color-magnitude diagrams developed in the last decade \citep{HarrZari01,Dolp02,JorgLind05,NaylJeff06}.

	The paper is organized in the following way. I start by briefly describing the problem. Then, I select a simplified case (clusters with fixed single age, metallicity, 
and extinction) and derive an analytical solution for the effects of SIMF uncertainties on individual cluster masses and the CMF for the case where uncertainties are small 
and masses are derived from $V$-band photometry (an appendix provides the mathematical formalism). Later, I use Monte Carlo simulations to analyze arbitrary mass 
uncertainties and develop an approximation using the Gamma distribution to obtain the behavior of the observed mass distribution for a given arbitrary cluster mass. The 
simulations are then extended to other photometric bands and known ages. Finally, the cases where the age or the age and the extinction are unknown and have to be derived 
from the same multi-band photometry as the mass are considered, and a method to obtain all of them simultaneously is presented and analyzed. I wrap up with a discussion on how these results
apply to real data and objects and a list of conclusions.

\section{Description of the problem}

\label{desprob}

	The standard technique to derive an observed cluster mass \mo\ with known age, extinction, distance, and metallicity
from spatially-integrated photometric data has two steps. First, the photometry is 
processed to account for distance and extinction to derive the luminosity in a given band (e.g. Johnson $V$), 
$L_V$. Second, evolutionary synthesis models are used to derive $Q_V$, the mass-to-luminosity ratio in that band for very large (real) cluster masses $m$, as a 
function of cluster age:

\begin{equation}
Q_V = \lim_{m\rightarrow \infty} \frac{m}{L_V}.
\end{equation}

	The observed cluster masses are then:

\begin{equation}
\mo = Q_V L_V.
\label{molv}
\end{equation}

	What is the uncertainty $\sigma$ associated to the measurement of \mo? There are three sources of uncertainty to consider:
photometric, model, and SIMF sampling. For CCD measurements, the photometric uncertainty arises in most cases from the Poisson statistics of 
the detected radiation (but note that in some cases read noise and background subtraction can also play an effect). Under model uncertainties I include a 
number of effects that originate in our transformation of the photometry to $L_V$ and the value of $Q_V$, such as an inadequate characterization of 
evolutionary tracks and stellar atmospheres, incorrect calculation of ages and extinctions, and errors in the zero points of the photometric systems. 
The third uncertainty source, SIMF sampling, arises because the SIMF (understood as a probability density function) is only well sampled at very large masses;
in other cases two clusters of the same mass can have significantly different populations of massive stars, which are the most scarce ones but can dominate 
the light output from young clusters. Therefore, SIMF sampling in a cluster dictates that two clusters of the same $L_V$, age, and metallicity do not necessarily have the same 
mass, an effect that becomes especially important for low-mass clusters.

	Which uncertainty source dominates? The ultimate answer will depend, of course, on the data and methods used to derive the observed masses. 
The behavior of model uncertainties is difficult to calculate because of the many parameters involved. However, given that photometric zero points are usually
known to within 2\% or better \citep{Maiz06a,Maiz07a}, that current evolutionary synthesis models such as Starburst 99 \citep{Leitetal99} are quite 
capable of accurately reproducing the integrated spectra of massive clusters of different ages, and that nowadays relatively sophisticated methods can be
used to compare photometric data to model predictions \citep{Andeetal04,Maiz04c}, it is safe to put a 5\% cap on model uncertainties\footnote{If SIMF sampling is not relevant. As we shall
see later on, some of the model uncertainties (e.g. extinction) can be coupled with SIMF sampling issues and, thus, be greatly enhanced.}. 

	To characterize the other two types of uncertainties (photometric and SIMF sampling), I first consider a simplified example that is 
relevant to current research: a family of single-age 
(10 Ma), solar metallicity, constant extinction ($A_V = 1.0$) stellar clusters located at the distance of the Antennae galaxies (22 Mpc, 
\citealt{Schwetal08}) observed with the Hubble Space Telescope ACS/WFC camera using the F550M filter with two exposures of 1250 s each. The photometric 
uncertainties were calculated using the ACS Imaging Exposure Time Calculator\footnote{\tt http://etc.stsci.edu/webetc/acsImagingETC.jsp}. The SIMF sampling 
uncertainties were calculated using the evolutionary synthesis add-on module available in version 3.1 of CHORIZOS \citep{Maiz04c}, which also produced the Monte 
Carlo simulations that are described later on. Results are shown in Table~\ref{unctypes}, with both a low-mass and a high-mass cluster as examples, and a 
clear conclusion can be extracted from them: at least in this simplified case\footnote{But likely also in many others, since for objects closer than the Antennae photometric
uncertainties should be smaller.} {\bf SIMF sampling is the largest source of uncertainty in the determination of the masses of 
unresolved stellar clusters} for masses lower than $10^5$ \Ms/ and is clearly dominant in the more interesting (and currently actively debated) range 
$10^2-10^4$ \Ms/. 

\section{An analytical solution for \fmo\ when $\sigma \ll m$}

	As a first step towards the characterization of the possible biases caused by mass uncertainties in stellar cluster measurements, I develop an
analytical solution that describes the behavior for the case $\sigma/m \ll 1$. I will consider that the real stellar cluster mass distribution, \fm, is given 
by a power law with slope $\gamma$ (see Eqn.~\ref{fm} and Appendix A for the Bayesian formalism used in this paper) and that $\sigma$ has a power-law dependence on mass:

\begin{equation}
\sigma(m) = \sr (m/\mr)^\beta,
\label{sigma1}
\end{equation}

\noindent where \mr\ is a reference mass, \sr\ the uncertainty for that reference mass, and $\beta$ a power-law exponent, with $\beta \le 1$. 
We then have that the joint distribution of observed and real masses (see Eqn~\ref{fxox}) is:

\begin{eqnarray}
\fmom & = & \frac{A}{\sqrt{2\pi}\sr(m/\mr)^\beta}\exp\left[-\frac{1}{2}\left(\frac{\mo-m}{\sr(m/\mr)^\beta}\right)^2\right] m^\gamma \\
      & = & \frac{A\,\mr^\beta\mo^{\gamma-\beta}}{\sqrt{2\pi}\sr}\exp\left[-\alpha^2y^2(1+y)^{-2\beta}/2\right](1+y)^{\gamma-\beta},
\end{eqnarray}

\noindent where I have defined:

\begin{eqnarray}
y      & = & (m-\mo)/\mo \\
\alpha & = & \mo^{1-\beta}\mr^\beta/\sr .
\end{eqnarray}

	In order to guarantee that $\sigma/m \ll 1$ I will consider only the range $\mo>\mr$ with the additional assumption that $\sr/\mr\ll 1$. In that
case, it is easy to see that $\alpha\gg 1$ and that \fmom\ is non-negligible only in those regions where $|\alpha y| \lesssim 1$, i.e. where $|y|\ll 1$ 
and the observed mass is not too different from the real one. Under such circumstances we can do a Taylor expansion of the terms in $(1+y)$ to obtain:

\begin{equation}
\begin{split}
\fmom \approx & \; \frac{A\,\mr^\beta\mo^{\gamma-\beta}}{\sqrt{2\pi}\sr}\exp\left(-\alpha^2y^2/2\right) \times            \\
              & \;      \left( 1 + (\gamma-\beta)y + \frac{(\gamma-\beta)(\gamma-\beta-1)}{2}y^2+                \right.  \\
              & \; \;\; \left.\beta\alpha^2y^3+\beta(\gamma-2\beta-1/2)\alpha^2y^4+\frac{1}{2}\beta^2\alpha^4y^6 \right), \\
\end{split}
\label{fmomanal}
\end{equation}

\noindent where I have kept all the terms up to $\alpha^ny^{n+2}$. Hence, the distribution of observed masses (see Eqn.~\ref{fxo}) is given by:

\begin{equation}
\fmo \approx \mo^\gamma \left(1 + \frac{(\gamma+2\beta)(\gamma+2\beta-1)}{2\alpha^2}\right).
\label{fmoanal}
\end{equation}

I can now use Eqns.~\ref{fmomanal}~and~\ref{fmoanal} to calculate first the mass correction for an individual cluster ($c$, see Eqn.~\ref{canal3}), and then 
the fractional number change ($p$, see Eqn.~\ref{panal3}) and the slope change (\dgamma, see Eqn.~\ref{dgammaanal3}) between the real and observed distributions
for the approximation described above:

\begin{equation}
c = \frac{\gamma+2\beta}{\alpha^2} = (\gamma+2\beta)\left(\frac{\sr}{\mr^\beta\mo^{1-\beta}}\right)^2,
\label{canal1}
\end{equation}

\begin{equation}
p = \frac{(\gamma+2\beta)(\gamma+2\beta-1)}{2\alpha^2} = \frac{(\gamma+2\beta)(\gamma+2\beta-1)}{2}\left(\frac{\sr}{\mr^\beta\mo^{1-\beta}}\right)^2,
\label{panal1}
\end{equation}

\begin{equation}
\dgamma = -\frac{(1-\beta)(\gamma+2\beta)(\gamma+2\beta-1)}{\alpha^2} = 
-(1-\beta)(\gamma+2\beta)(\gamma+2\beta-1)\left(\frac{\sr}{\mr^\beta\mo^{1-\beta}}\right)^2.
\label{dgammaanal1}
\end{equation}

	Equations~\ref{canal1},~\ref{panal1},~and~\ref{dgammaanal1} reveal that, to first order, the effect of uncertainties on $c$, $p$, and \dgamma\ depend
on \mo\ as $\mo^{-2(1-\beta)}$ so that they tend to zero for large masses unless $\beta = 1$. The sign and specific
magnitude of the effect of uncertainties on the three quantities depend on the specific values of $\gamma$ and $\beta$. It is interesting to note that
for $\gamma = -2\beta$, $c$, $p$, and \dgamma\ are all zero, indicating that the observed and real distributions are identical and that no overall shift
of the masses is observed to this degree of approximation. For $\gamma = 1-2\beta$, $c > 0$ but $p =\dgamma = 0$, indicating that even though the two 
distributions are identical there is an overall shift from higher to lower masses for a given value i.e. real masses are larger than observed ones on 
average\footnote{Of course, this is possible because I am not considering effects near the edges of \fm.}. For $\beta =1$, Eqn.~\ref{dgammaanal1} 
indicates that the observed distribution is always a power law with the same slope as the real one and Eqns.~\ref{canal1}~and~\ref{panal1} give values
of $c$ and $p$ independent of \mo. 

	When uncertainties originate from SIMF sampling ($\beta = 0.5$), Eqns.~\ref{canal1},~\ref{panal1},~and~\ref{dgammaanal1} give:

\begin{equation}
c = \frac{\gamma+1}{\alpha^2} = \frac{(\gamma+1)\sr^2}{\mr\mo},
\label{canal2}
\end{equation}

\begin{equation}
p = \frac{\gamma(\gamma+1)}{2\alpha^2} = \frac{\gamma(\gamma+1)\sr^2}{2\mr\mo},
\label{panal2}
\end{equation}

\begin{equation}
\dgamma = -\frac{\gamma(\gamma+1)}{2\alpha^2} = -\frac{\gamma(\gamma+1)\sr^2}{2\mr\mo}.
\label{dgammaanal2}
\end{equation}

	Therefore, in the low-uncertainty limit SIMF sampling generates mass corrections for individual clusters, fractional number changes, and CMF slope changes 
for large masses that are inversely proportional to the observed masses. For $\gamma < -1$, $c$ and \dgamma\ are negative and $p$ is positive.

\section{Monte Carlo simulations for \fmocm}

	The analytical solution developed in the previous section can help us characterize the effect of uncertainties on the observed mass distribution
for stellar clusters but, considered by itself, it has two shortcomings. The first one is the scale of the effect e.g. if we rewrite Eqn.~\ref{sigma1} for
$\beta=0.5$ as:

\begin{equation}
\sigma = \sqrt{\theta\, m},
\label{sigma2}
\end{equation}

\noindent we still need to calculate the value of $\theta$ for the case of interest. For a Poisson process, Eqn.~\ref{sigma2} is satisfied, so I will call $\theta$ the Poisson mass. 
$\theta$ can be thought of as characteristic cluster mass since for $m = \theta$, $\sigma = m$, and will be used extensively in this paper as a measurement of the stochasticity 
of clusters as a function of age and other parameters. The choice of notation will become clear when I introduce the Gamma approximation in the next subsection.
The second shortcoming is that the analytical solution is expected to 
work only when the effect of uncertainties is relatively small and when we are using a power law without mass limits (which, of course, is unphysical because such 
distribution cannot be normalized). In order to evaluate its range of validity and to consider a more realistic case with possible large values of 
$\sigma/m$ and mass limits, we need to go beyond an analytical approach and use Monte Carlo simulations of \fmocm. Such simulations can also provide us with 
the value of $\theta$ and test whether Eqn.~\ref{sigma2} is indeed correct. In this section and the following one I develop an example that, though a 
simplification of a real case, retains its most important effects. Later on I include some of the additional complications related to real data.

	Since I have already established the age, metallicity, extinction, and distance of our example in section~\ref{desprob}, the next parameter that 
needs to be determined is $\gamma$. A number of independent results indicate that the CMF has a slope close to $-2.0$ in many circumstances (e.g. 
\citealt{ElsoFall85b,ZhanFall99,Oeyetal04}), so I fix $\gamma = -2.0$ for the simulations. For that value of $\gamma$ the analytical approximation gives 
$c = -1/\alpha^2$, $p = 1/\alpha^2$, and $\dgamma = -1/\alpha^2$. I also fix $\mu = 10^6$ \Ms/ and let \ml\ vary between 1 and 1000 \Ms/.

	The next issue to consider is the type of Monte Carlo simulation. Simulations of stellar clusters typically use either a fixed number 
of stars or a fixed total stellar mass \citep{Cervetal02b,WeidKrou06}. Here our goal is to generate \fmocm, so I use Monte Carlo simulations of the 
second type. We also need to select the photometric band to calculate the observed mass (Eqn.~\ref{molv}): our choice here is the most popular one, Johnson
$V$. To generate the synthetic clusters I use the evolutionary synthesis code available as an add-on to the latest version of CHORIZOS \citep{Maiz04c}. 
Each simulated 10 Ma cluster is a random realization of a Kroupa IMF \citep{Krou01} between 0.1 and 120
\Ms/ (note that it is possible that some stars may have already exploded as SNe) using a Padova isochrone \citep{Marietal08} for $m < 7$ \Ms/ and a Geneva 
isochrone \citep{LejeScha01} for $m > 7$ \Ms/. The massive end of such an isochrone corresponds to a red supergiant with an initial mass of 18.29 \Ms/ and
$M_V = -6.13$. The observed mass associated with the lowest luminosity limit \moLLL\ (i.e. the maximum observed mass that can be produced by a ``single-star 
cluster'' of that age, see \citealt{CervLuri04} who use $\mathcal{M}^{\rm min}$ for such a quantity) for such a cluster is 2795 \Ms/, with the 
``single-star cluster'' being in reality a 16.94 \Ms/ blue supergiant with $M_V = -7.47$.

\begin{figure}
\centerline{\includegraphics*[width=\linewidth, bb=130 400 475 720]{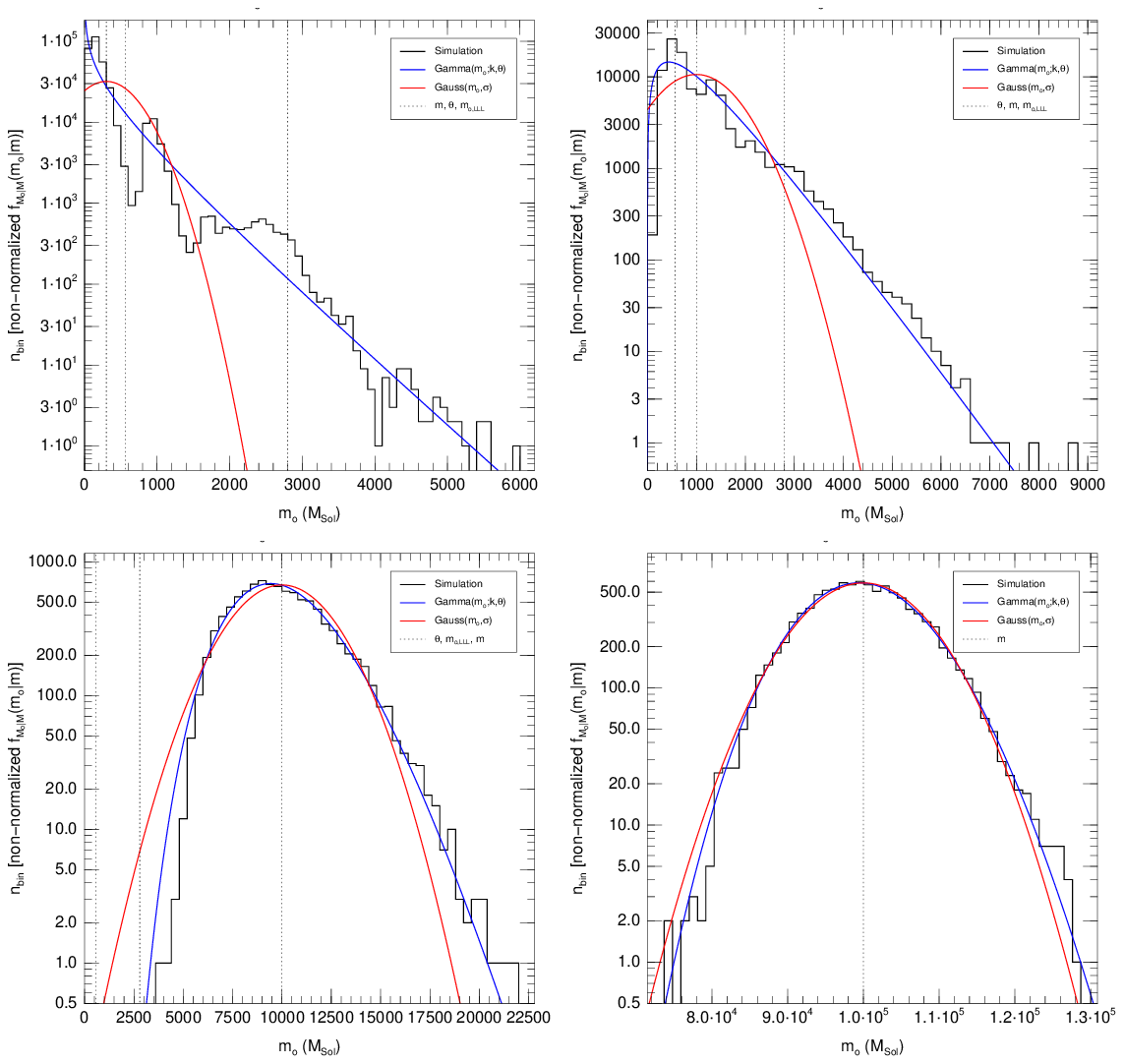}}
\caption{Results of the Monte Carlo simulations of \fmocm\ and Gaussian and Gamma fits for 10 Ma clusters observed in $V$ with four real masses: 300 \Ms/ 
(upper left), 1000 \Ms/ (upper right), 10\,000 \Ms/ (lower left), and 100\,000 \Ms/ (lower right). The vertical dotted lines show the values of the real mass 
$m$ and, when visible, of $\theta$ and \moLLL.}
\label{fmocm1}
\end{figure}

	I generate Monte Carlo simulations of \fmocm\ for eight values of $m$ between \sci{3}{2} \Ms/ and \sci{1}{6} \Ms/. \ns, the number of simulations, was adjusted for 
each of the eight values of $m$ to guarantee a good sampling of \fmocm. The output is shown in 
Table~\ref{MonteCarlores} and in Fig.~\ref{fmocm1}. 

	The first result is that \mo\ indeed behaves as a Poisson-like variable, with an average value $\overline{m}_{\rm o}$ 
almost identical to $m$ and a standard deviation $\sigma$ that satisfies Eqn.~\ref{sigma2} to a high degree, as evidenced by the near-constant value of 
$\theta$ derived for different masses. From Table~\ref{MonteCarlores} I obtain an average value\footnote{This value, of course, is only applicable to the 
type of cluster and mass-measurement method described here. In other circumstances $\theta$ should be different, as we shall see later on.} for $\theta$ of 
566 \Ms/, which will be used subsequently.

	The second result is that, for low values of $m$, \fmocm\ highly deviates from a Gaussian (see top plots of Fig.~\ref{fmocm1}). Such a
phenomenon has been described by \citet{CervLuri06} and it was easily predictable (but too often neglected in the literature): 
as $m$ decreases, $\sigma/m$ increases, with $\sigma = m$ for
$m = \theta$ and $\sigma > m$ for lower values of $m$. Under such circumstances a Gaussian cannot provide an accurate description because it produces
a significant fraction of stars with negative masses, which is obviously unphysical. Furthermore, the complex shape of the luminosity function produced by
the existence of fast post-MS evolutionary phases (e.g. yellow supergiants, see \citealt{CervLuri06}) induces the existence of complex structures in 
\fmocm\ for low values of $m$ (e.g. the peak seen around 1000 \Ms/ in the upper left plot of Fig.~\ref{fmocm1}, which is real and not the result of using a
small number for \ns). For higher real masses a Gaussian produces a reasonable approximation of \fmocm, as evidenced in the bottom plots of Fig.~\ref{fmocm1}.

	The non-Gaussianity of \fmocm\ introduces a problem. Eqns.~\ref{fxocx},~\ref{fxox},~and~\ref{fxo} are no longer valid and since our goal is to 
calculate \fmo, defined as:

\begin{equation}
\fmo = \int \fmom \, dm = \int \fmocm \fm \, dm,
\label{fmo}
\end{equation}

\noindent we need to be able to characterize \fmocm\ in a fine grid, i.e. to compute \fmocm\ for an arbitrary value of $m$ within a reasonable amount of
computing time. Monte Carlo simulations are computationally expensive (the time needed to generate the eight simulations in Table~\ref{MonteCarlores} 
was several hours) and generating hundreds or thousands of them may be simply prohibitive, so an alternative is needed. This is dealt with in the next
section.

\section{The Gamma approximation and a solution for \fmo}

	One approach to the generation of \fmocm\ in a fine grid is that of \citet{CervLuri06}. Those authors suggest going beyond the Gaussian approximation
by using an Edgeworth series that provides the correct values for the skewness and kurtosis of the distribution function. 
An alternative approach is to find a family of distributions that has the desirable properties of: (a) having the variance equal to the square of the
mean; (b) being zero for negative values of the observed mass; (c) reasonably approximating the skewness, kurtosis, and asymptotic behavior for $\mo \rightarrow 0$ and 
$\mo \rightarrow \infty$ of the Monte Carlo simulations; and (d) behaving like a Gaussian in the limit $m\rightarrow \infty$. One choice is the Gamma
distribution, a two-parameter distribution defined by the probability density function:

\begin{equation}
f(x; k, \theta) = x^{k-1}\frac{e^{-x/\theta}}{\theta^k \, \Gamma(k)},
\end{equation}

\noindent where $k$ and $\theta$ are the scale and shape parameters (both positive), respectively, and $\Gamma(k)$ is the normalization factor:

\begin{equation}
\Gamma(k) = \int_0^\infty t^{k-1} e^{-t} dt.
\end{equation}

	The Gamma distribution has mean $k\theta$ and variance $k\theta^2$. Comparing this with the desired properties for \fmocm, I find that
$\theta$ has to be defined as in Eqn.~\ref{sigma2} and that $k=m/\theta$. Hence, the Gamma-like distribution function approximation for \fmocm\
with the appropriate mean and variance can be written as:

\begin{equation}
\fmocm = \mo^{m/\theta-1}\frac{e^{-m_{\rm o}/\theta}}{\theta^{m/\theta} \, \Gamma(m/\theta)},
\label{fmocm2}
\end{equation}

	This expression is easily computable and should be understood as a function of \mo, with $m$ and $\theta$ as parameters (i.e.
it is a distribution of observed masses for given real masses and Poisson mass $\theta$). How good is this Gamma approximation? Figure~\ref{fmocm1} shows
it for four different values of $m$ and Table~\ref{MonteCarlores} gives the values for $k$ ($= m/\theta$, with $\theta$ being the average value previously 
calculated), \sk\ (the skewness from the Monte Carlo simulations), \skg\ (the skewness from the Gamma approximation), \ku\ (the kurtosis from the Monte Carlo 
simulations), and \kug\ (the kurtosis from the Gamma approximation). It can be seen that Eqn.~\ref{fmocm2} satisfies the previously stated desirable
properties, only underestimating the skewness by $\approx 25\%$ and the kurtosis by $\approx 35\%$. Also, since the Gamma distribution
is a smooth, single-peaked distribution it cannot reproduce the complex structures seen in the Monte Carlo simulations for low values of $m$ but note that 
those peaks are not centered at the same values of \mo\ for different values of $m$ (e.g. compare the region around 1000 \Ms/ for the two top plots of
Fig.~\ref{fmocm1}). Therefore, when one integrates over $m$ in Eqn.~\ref{fmocm2} those small deviations from the gamma approximation should be mostly smoothed
out. On the other hand, the asymptotic behavior for small and large values of \mo\ is excellent, and this is important in order to correctly characterize how 
many low-mass (e.g $\sim 300$ \Ms/) clusters are masquerading as intermediate-mass (e.g. 3000 \Ms/) clusters through sampling effects. 

\begin{figure}
\centerline{\includegraphics*[width=0.62\linewidth, bb=160 190 445 720]{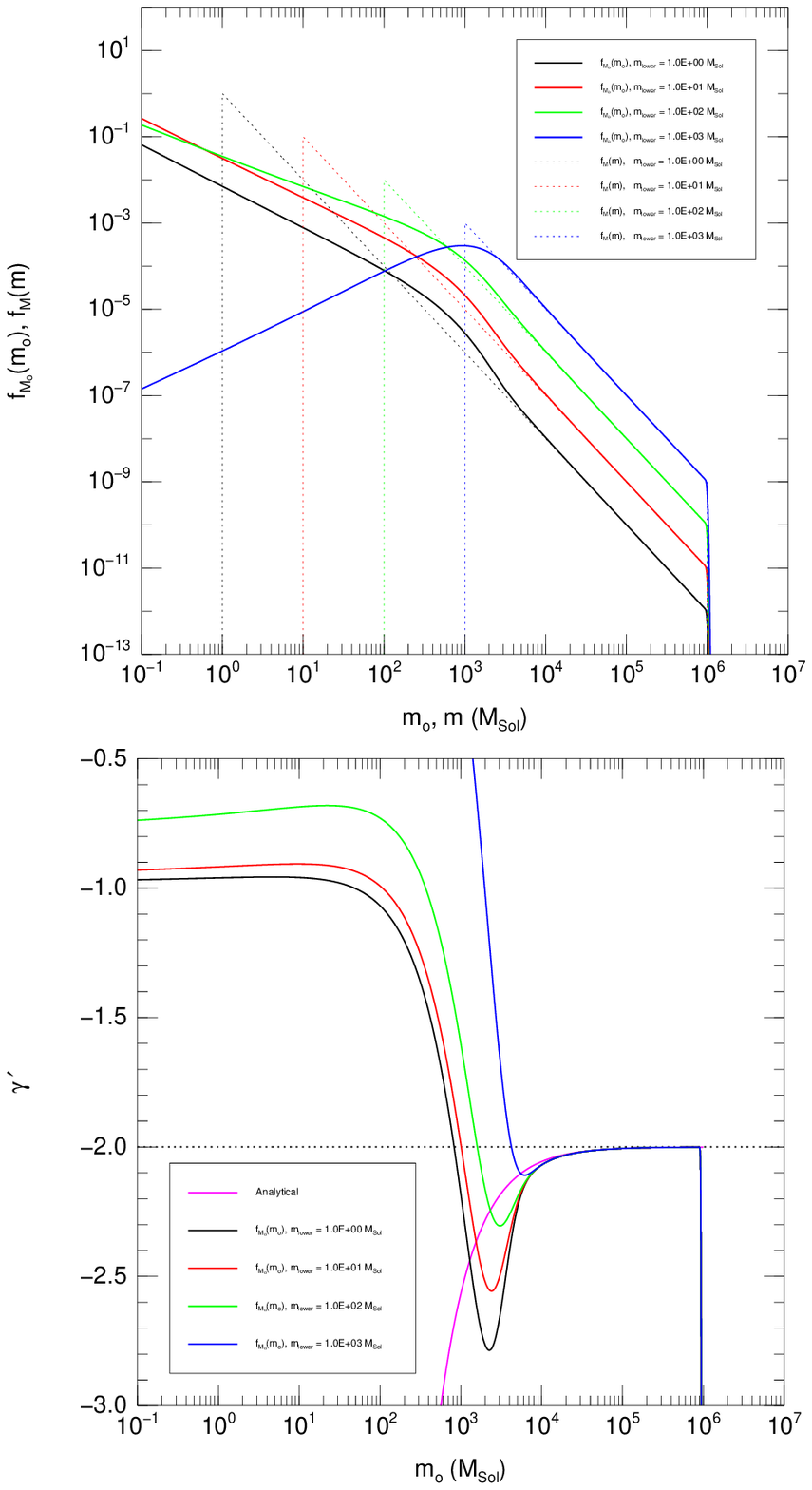}}
\caption{Observed [\fmo, continuous lines] and real [\fm, dotted lines] mass distributions (top) and slopes (\gammap, bottom) using the Gamma 
approximation with 10 Ma clusters observed with $V$ ($\theta = 566$ \Ms/).}
\label{fmofm1}
\end{figure}

	Using the Gamma approximation for \fmocm, I apply Eqn.~\ref{fmo} to generate \fmo\ for four values of \ml\ between 1 and 1000 \Ms/ (note that 
1~\Ms/ and 1000~\Ms/ themselves are rather unrealistic extremes for \ml, while 10~\Ms/ and 100~\Ms/ are more reasonable values). Results
are presented in Fig.~\ref{fmofm1}, which show the observed and real distributions (top) and the slope of the observed distributions along with the
analytical solution. As expected, the analytical solution is a good approximation for $\mo \gtrsim 10^4$~\Ms/, since $\sigma/m = 0.24$ for $m = 10^4$~\Ms/, but fails 
for low-mass clusters.

	Figure~\ref{fmofm1} shows that \fmo\ is relatively similar to \fm\ for $\mo \gtrsim 10^4$~\Ms/ (as expected from the analytical solution) but that
there are significant differences for lower observed masses. First, in all cases except $\ml = 1000$~\Ms/ there is a ``hump'' (overdensity of \fmo\ with 
respect to \fm) at $\sim 1000$~\Ms/ which is generated by lower-mass clusters with relatively large apparent masses (due to the existence of one or several 
very bright stars). The height of the ``hump'' increases as \ml\ decreases. Second, to the left of
the hump \fmo\ asymptotically tends to a power law with $\gammap > -1.0$, with \gammap\ decreasing with \ml. Third, in order to connect the two asymptotic
behaviors ($\gammap > -1.0$ for low observed masses and $\dgamma = -\theta/\mo$ for high observed masses) it is necessary for the observed slope \gammap\ 
to have a minimum with a value below $-2.0$ at a few thousand solar masses. As seen in Fig.~\ref{fmofm1}, the depth of that minimum increases with decreasing \ml. 

\section{Filter, age, and extinction effects}

	The case analyzed in the previous sections is quite simplified since we are using one single photometric band and we assume fixed known values for the age, 
metallicity, extinction, and distance to the clusters. In a more realistic case, one uses two or more filters and has to determine some or all of those four
observables at the same time as 
the mass. It is easy to envision a case where all those quantities are variable but, in order not to excessively complicate the analysis, I will assume 
that the distance is known and the metallicity is fixed to solar, so the only possible variables are age and extinction. Also, only a uniform foreground screen far away 
from the cluster will be used to model the extinction.

\subsection{Other single filters and known ages}

\begin{figure}
\centerline{\includegraphics*[width=\linewidth]{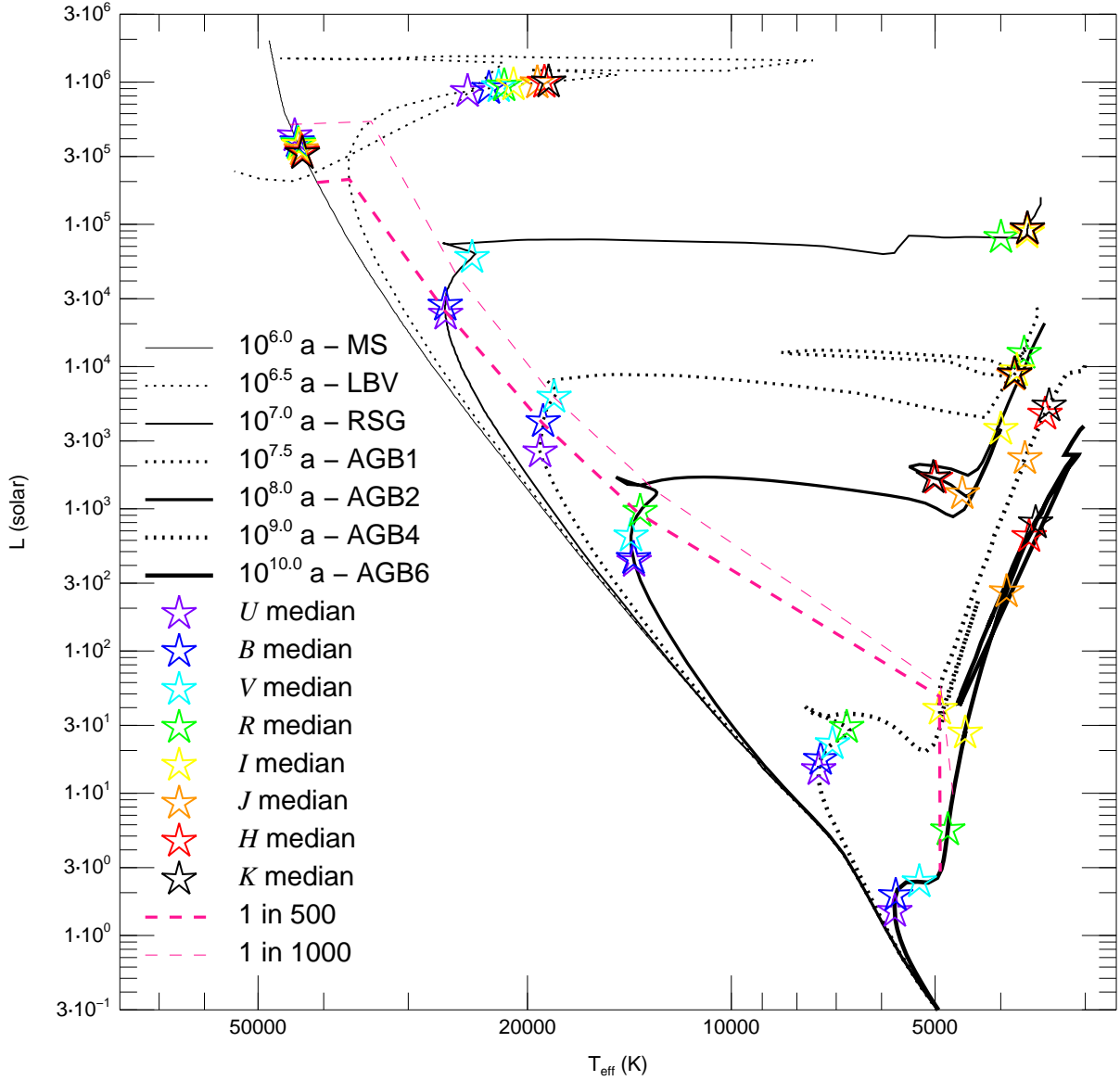}}
\caption{Isochrones for seven of the nine ages used in this paper. For each isochrone, symbols of different colors are used to indicate the median mass for the luminosity in
the eight filters $U$, $B$, $V$, $R$, $I$, $J$, $H$, and $K$ assuming a Kroupa IMF between 0.1 \Ms/ and 120 \Ms/. The two dashed lines join the points in each isochrone
beyond which 0.20\% and 0.10\%, respectively, of the remaining stars in a well-sampled Kroupa IMF are located. The two remaining isochrones are not included for the sake
of clarity.}
\label{isoc}
\end{figure}

	In order to extend the analysis beyond the results for 10 Ma clusters using $V$ derived in the previous section, I start by repeating the Monte Carlo 
simulations of \fmocm\ for seven additional filters (Johnson $U$ and $B$, Cousins $R$ and $I$, and 2MASS $J$, $H$, and $K$) and
eight additional ages, 1 Ma, 3.16 Ma, 31.6 Ma, 100 Ma, 316 Ma, 1 Ga, 3.16 Ga, and 10 Ga (i.e. logarithmic values of 6.0, 6.5, 7.5, 8.0, 8.5, 9.0, 9.5, and 10.0 years, 
respectively), all of them with solar metallicity. The isochrones were derived from the
same sources as the 10 Ma one and the throughputs and zero points for the filters are those described in \citet{Maiz07a}. The most luminous stars for the
first three isochrones are main-sequence stars (MS, 1 Ma), luminous blue variables (LBV, 3.16 Ma)\footnote{Several points need to be mentioned regarding the 3.16 Ma case. 
First, the isochrone published by \citet{LejeScha01} stops near the point where stars become LBVs and does not include more evolved (WR) stars. I address this point 
by interpolating the associated evolutionary tracks to extend the isochrone to the end of the Wolf-Rayet phase. Second, the evolution near the LBV stage is quite fast and, to
some degree, unknown, with the star being capable of rapidly changing its surface temperature and switching between BSG, ASG, and WR spectral types even within a human 
lifetime (see \citealt{Walbetal08} for an example). Third, it is not even clear whether all stars survive the LBV stage to become WRs or instead explode as SNe while they
are LBVs \citep{Smitetal07b}. In any case, I verified that none of the above fundamentally affects the results in this section by running comparison simulations with the 
post-LBV phases excluded.}, and red supergiants (RSG, 10 Ma), respectively, while for the last six (between 31.6 Ma and 10 Ga) they are asymptotic giant branch stars (AGB). 
Accordingly, I will refer to each of the nine ages as the MS, LBV, RSG, AGB1, AGB2, AGB3, AGB4, AGB5, and AGB6 phases, respectively\footnote{Note that the phases 
are named after the most luminous star possible for each age, not after the most luminous star that exists in a given cluster (i.e. a realization of the SIMF for that age).
For example, a low-mass 10 Ma cluster can have only main-sequence stars because it does not have stars of the right mass (close to 18 \Ms/) to have a RSG at that precise 
age.}. Figure~\ref{isoc} shows the isochrones for seven of the nine ages.

	For all filters and ages, I find that \mo\ behaves as a Poisson variable and that the Gamma distribution is a good approximation to \fmocm\ over a large range of 
real masses. The values of $\theta$ (Table~\ref{theta}) show a large spread between 23 \Ms/ and 5717 \Ms/, with two clear patterns: [1] As a function of
age, $\theta$ starts being small in the MS phase, grows very rapidly when stars start evolving and the cluster enters the LBV phase, reaches a maximum either in the 
LBV (optical bands) or RSG (NIR bands) phase, decreases until a mid-AGB phase and then increases again until the last age analyzed. [2] With the only exception of the MS phase, 
$\theta$ increases as a a function of the effective wavelength, with the effect becoming larger with age. The trend in wavelength is reversed in the initial MS phase but the 
effect there is relative small ($\theta_K/\theta_U = 0.6$ at 1~Ma but $247$ at 1~Ga).

	What is the reason for the behavior of $\theta$ as a function of age and wavelength? One way to analyze the problem is to compute for a given isochrone and photometric 
band the product of the SIMF and the luminosity in that filter and to obtain the stellar mass at which the median of such a distribution takes place. That median luminosity point for a 
given filter tells us the location in the isochrone beyond which 50\% of the light in that filter is produced. Then, to decide whether SIMF sampling effects are expected to be relevant
in that band for a cluster with a given number of remaining stars we should compare that location to the point beyond which one expects only a small fraction of the stars to be present 
(i.e. the mass where the integral of the SIMF from there to the largest surviving mass is a small fraction of the integral for all the surviving masses). Such a comparison can be seen in 
Fig.~\ref{isoc}, which shows for the median luminosity points and the locations beyond where the fraction of surviving stars is 1/500 (0.2\% of the stars) and 1/1000 (0.1\% of the 
stars).

	The behavior of $\theta$ is now easily understood. For the MS phase, the median luminosity points for all filters are clustered between the 1/500 and 1/1000 points, with the value 
increasing towards shortest wavelengths. Hence, I would expect relatively low values of $\theta$, especially for the NIR bands. As we move to the LBV phase, two important changes take place: 
the most massive stars significantly increase their luminosity (with the rest of the stars remaining basically unchanged) and their surface temperatures switch from increasing to decreasing
with mass\footnote{Excluding the post-LBV phase, which contains a very low fraction of the stars and of generally lower luminosity.}. 
This causes the median luminosity points to rise beyond the 1/1000 point (producing a large increase in $\theta$) and their order as a function of wavelength to reverse
(inverting the tendency of $\theta$ with wavelength). Nevertheless, in both the MS and the LBV phase the wavelength dependence is relatively small because all median luminosity points are clustered together. 
This changes in the RSG phase, where the large spread in temperatures for the brightest surviving stars causes the median luminosity points to fall in very different points of the isochrone.  
The $U$ and $B$ bands fall well before the 1/500 point (thus returning to the low values of the MS phase) while the $R$, $I$, and NIR bands remain at the right hand of the isochrone,
well beyond the 1/1000 point. As we progress from 10 Ma to 100 Ma, the median luminosities move left and down (i.e. towards less evolved stellar stages)
with respect to equivalent points in the isochrone. This happens because the temperature difference between ```blue'' and ``red'' stars decreases as a function of age while the luminosity difference 
between RG/AGB and MS stars is kept moderately low and causes all the values of $\theta$ to decrease as a function of age. For older clusters the trend in $\theta$ is reversed because of the large
luminosity difference between evolved and MS stars. The effect especially important in the NIR since in the optical it is partially alleviated by the displacement of the 1/500 and 1/1000 points
towards the beginning of the RG part of the isochrone.

	One practical problem with $\theta$ is that Monte Carlo simulations are required to calculate it. Is there a way to avoid such simulations and to obtain $\theta$ (or at least
an estimate) directly from the isochrone and the SIMF? Two quantities that can be easily calculated and that quantify the stochasticity of \mo\ are \moLLL\ \citep{CervLuri04} and 
\neff\ \citep{Buzz89,Cervetal02b}, the latter defined as the effective fraction of stars contributing to the luminosity in a given band $X$:

\begin{equation}
\neff = \frac{\left(\int_{m_{\rm min}}^{m_{\rm max}} {\rm SIMF}(m) L_X(m) dm\right)^2}{\int_{m_{\rm min}}^{m_{\rm max}} {\rm SIMF}(m) L_X^2(m) dm}.
\label{neff}
\end{equation}

\begin{figure}
\centerline{\includegraphics*[width=\linewidth]{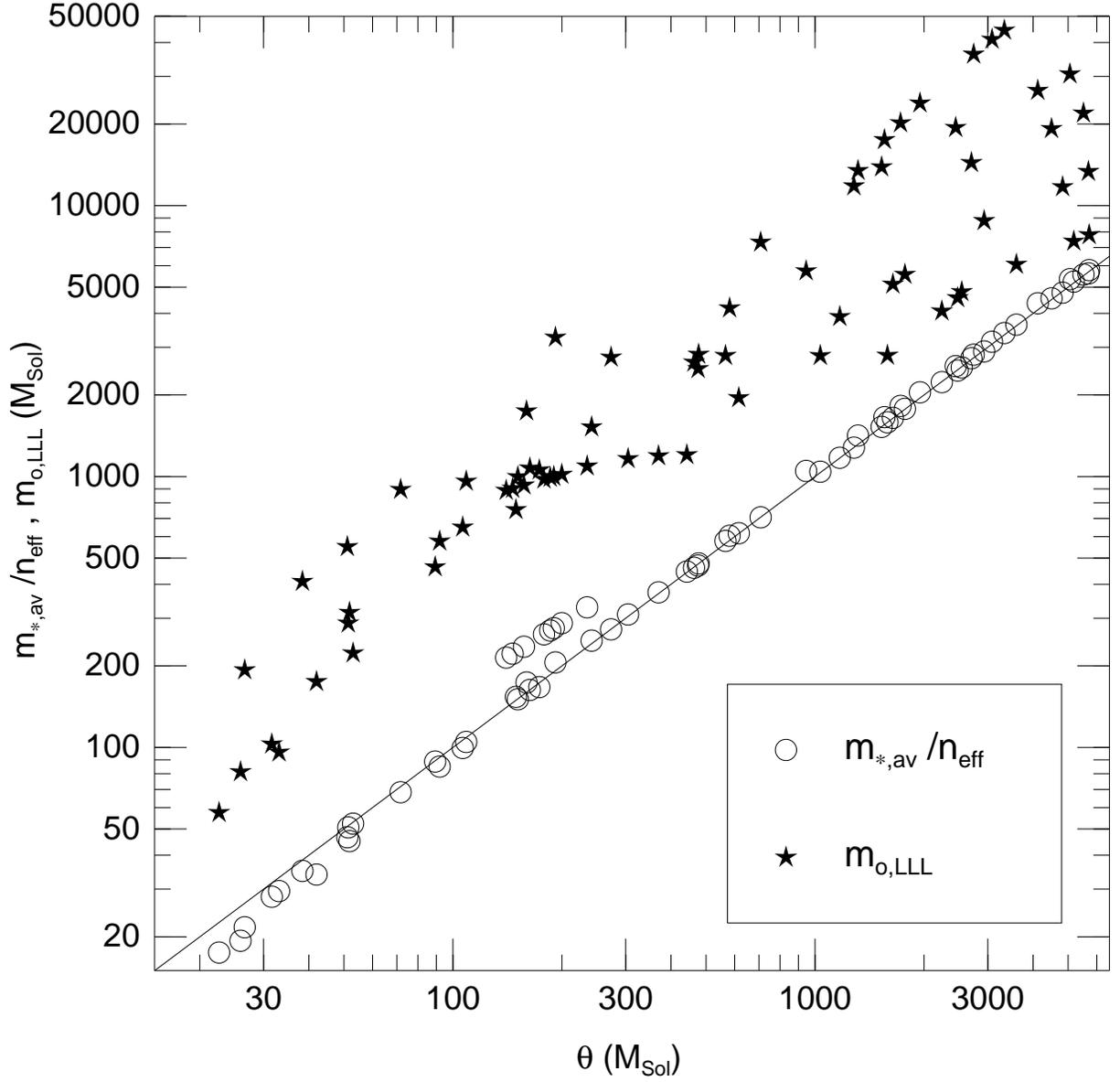}}
\caption{$m_{\rm \ast,av}$/\neff\ and \moLLL\ as a function of $\theta$ for the 72 Monte Carlo simulations (9 ages times 8 filters). The straight line marks the 1 to 1 locus.}
\label{thetaplot}
\end{figure}

	From Eqn.~\ref{sigma2}, one expects $\theta$ to be approximately \mstav/\neff, where \mstav\ is the average stellar mass for the SIMF (in our case, 0.71 \Ms/). I show in 
Fig.~\ref{thetaplot} the correspondence between, $\theta$, \moLLL, and \mstav/\neff\ for the 72 Monte Carlo simulations in this subsection. \moLLL\ is indeed correlated with $\theta$ but 
with a quite significant spread. This can also be seen in Tables~\ref{theta}~and~\ref{moLLL}, where the NIR values of $\theta$ for 3.16 Ma and 10 Ma are similar but those of \moLLL\ differ by an order of
magnitude.  On the other hand, \mstav/\neff\ is a much better predictor of the value of $\theta$. For the 72 values in Fig.~\ref{thetaplot}, the ratio of \mstav/\neff\ to $\theta$ has a mean of 1.04 
and a standard deviation of 0.17. Therefore, if all that is needed is a rough estimate of $\theta$, \mstav/\neff\ seems to do a reasonable job.

\begin{figure}
\centerline{\includegraphics*[width=0.62\linewidth, bb=160 190 445 720]{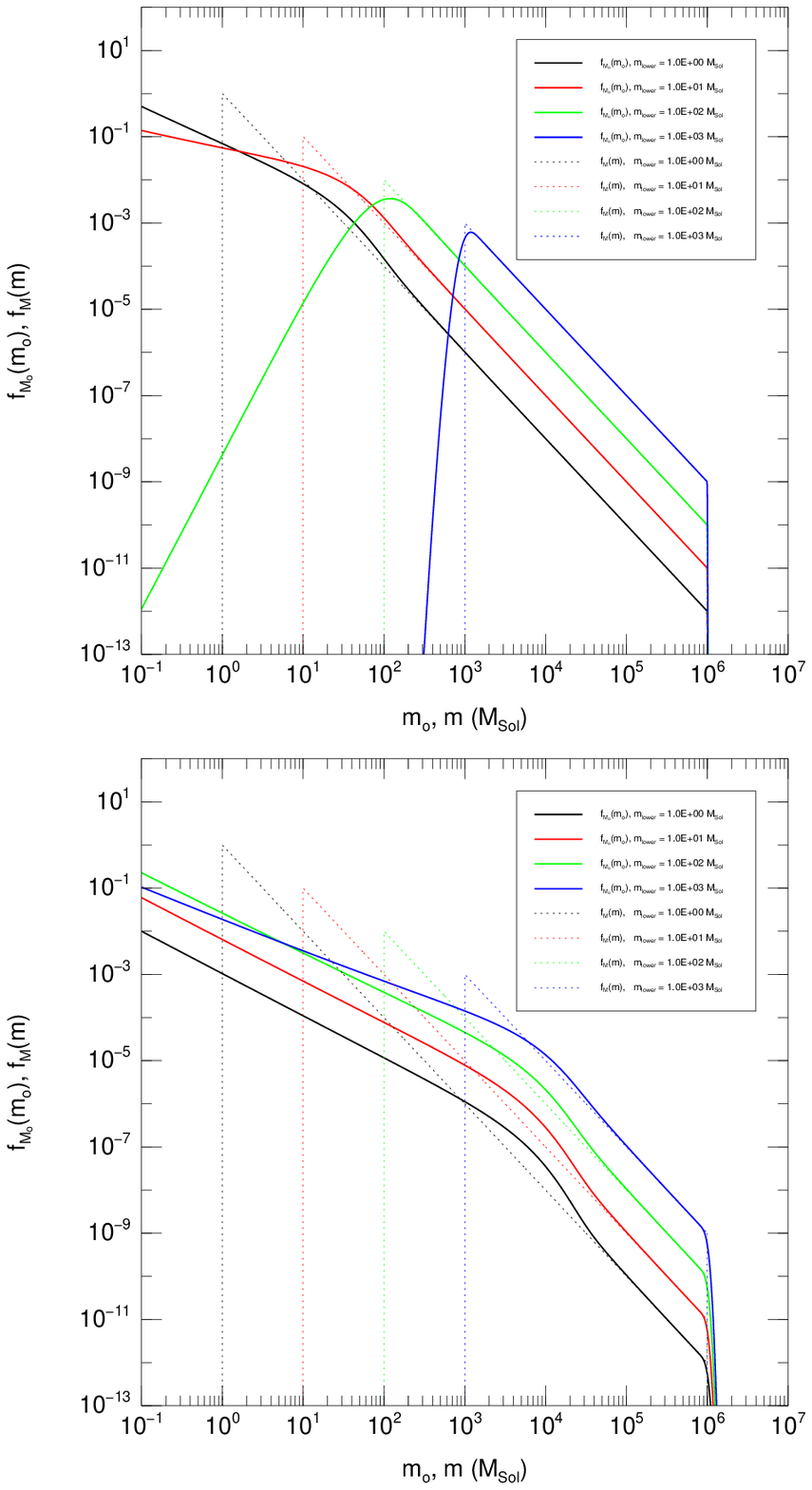}}
\caption{Observed [\fmo, continuous lines] and real [\fm, dotted lines] mass distributions using the Gamma 
approximation with [a] 1 Ga clusters observed with $B$ ($\theta = 23$ \Ms/, top) and [b] 316 Ma clusters observed with $K$ ($\theta = 5717$ \Ms/, bottom).}
\label{fmofm2}
\end{figure}

	With the $\theta$ values in hand and using the Gamma approximation for \fmocm, I can now apply Eqn.~\ref{fmo} to generate \fmo\ for clusters of different ages observed with
different filters.  In order to explore the full range of effects in \fmo\ I show the two extreme cases of $\theta$ in Table~\ref{theta} (23 \Ms/ and 5717 \Ms/) in Fig.~\ref{fmofm2} for 
the same values of \ml\ and \mu\ previously used. The overall structure seen in Fig.~\ref{fmofm2} is similar to that of the top panel of Fig.~\ref{fmofm1} but with some significant
differences. First, the ``hump'' changes position, with its central position located close to $2\theta$. Second, for fixed values of \ml\ and \mu\ the height of the ``hump'' increases with
$\theta$. Third, the combination of the dependences of the ``hump'' height with \ml\ (previously noted) and with $\theta$ makes it disappear for low values of $\theta$ not only for 
$\ml = 1000$ \Ms/ but also for $\ml = 100$ \Ms/. On the other hand, the ``hump'' is visible for all values of \ml\ for $\theta = 5717$ \Ms/. Finally, to the left of the ``hump'' the value of 
\gammap\ decreases as $\theta$ increases. In all cases I find that the analytical solution provides the correct asymptotic solution for large values. Therefore, if one wants to fix a value 
for \dgamma\ ($\ll 1$) as a tolerance limit for the effects of SIMF, it is possible to use Eqns.~\ref{sigma1}~and~\ref{sigma2} to obtain that the observed masses have to satisfy:

\begin{equation}
\mo > \frac{\gamma(\gamma+1)}{2\dgamma}\theta.
\label{morange}
\end{equation}

	For example, if we want to tolerate slope changes up to 0.1 and $\gamma = -2$, then $\mo > 10\,\theta$.

\subsection{Unknown ages}

	So far, I have considered that the cluster ages were known prior to the measurement of their masses. In most cases, this is unrealistic since one typically uses the photometry
to simultaneously determine mass and age. This is done by obtaining multi-band photometry, using the colors to first determine the age (and maybe also the extinction, which will be assumed
to be non-existent in this subsection), and then using one of the magnitudes to calculate the cluster mass. A classical example is that of \citet{ElsoFall85a}, who used spatially-integrated $UBV$ 
photometry to assign ages to a collection of LMC clusters by plotting their location in a $U-B$ versus $B-V$ diagram. 

	Working with color-color diagrams (e.g. $U-B$ versus $B-V$) to calculate ages has the limitation that only two photometric quantities can be used simultaneously. In order to take full advantage of 
the photometric information, one can use a multi-band $\chi^2$ minimization code such as CHORIZOS \citep{Maiz04c} or AnalySED \citep{Andeetal04}. Such codes allow one to explore a multi-dimensional
parameter space while using an arbitrarily large number of photometric bands. Increasing the number of filters has the advantage of using additional information that can be especially useful when the
observed photometry is similar to the model spectral energy distributions (SEDs) but not identical, such as when SIMF sampling effects are present. Thus, the additional filters can in some cases
(at least partially) ``correct'' the effect of one magnitude that is offset by such effects and provide a solution which is closer to the real one. Another advantage of a multi-band $\chi^2$ 
minimization code are the possibilities of assigning different weights to each filter and of estimating uncertainties in the output parameters.

	In this paper I will use CHORIZOS, a code in which several changes have been included (its latest version is 3.1) since it was described in \citet{Maiz04c}. CHORIZOS is now a full Bayesian code that
allows the use of priors and that can work with either magnitudes, colors, or spectrophotometry. It also now has a unified atmosphere grid that combines recent SEDs from different authors. 
More specifically, for O and B stars it uses the TLUSTY model atmospheres of \citet{LanzHube03,LanzHube07} and for late-type supergiants the MARCS model atmospheres of \citet{Plez08}. Version
3.1 of CHORIZOS has the added flexibility of dealing with almost any combination of up to 5 parameters and has an evolutionary synthesis add-on module. 

\begin{figure}
\centerline{\includegraphics*[width=0.59\linewidth, bb=160 180 445 720]{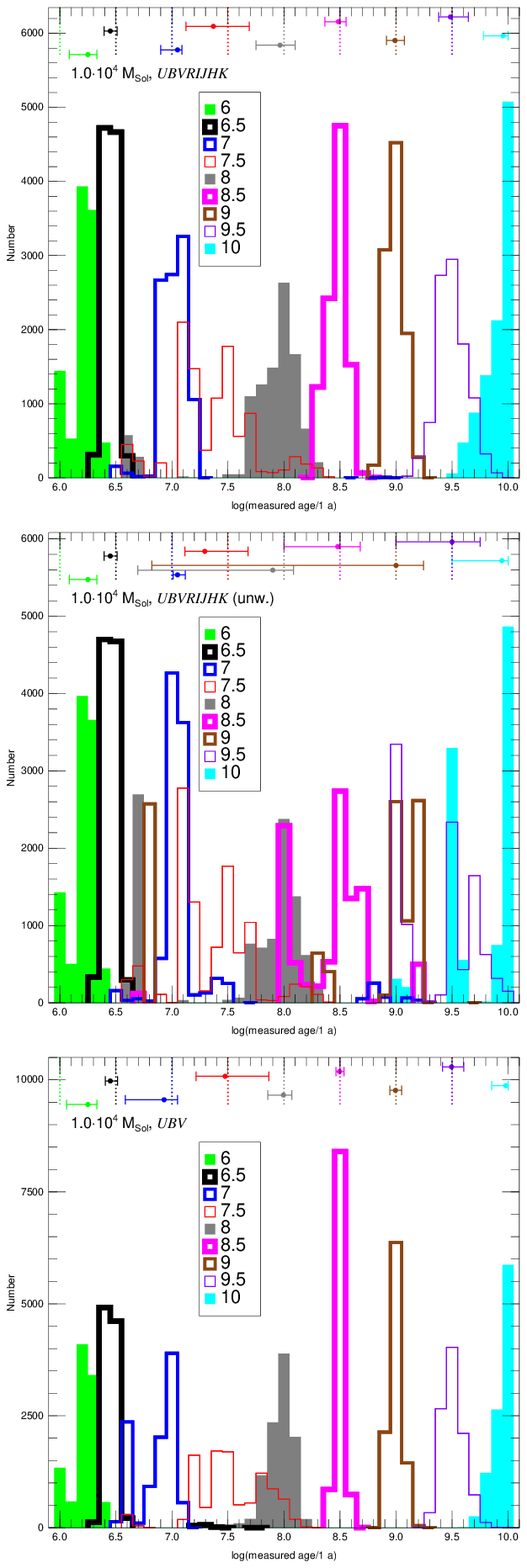}}
\caption{Distribution of observed ages for the nine ages and three execution types described in the text for clusters of $10^4$ \Ms/ and known extinction. At the top of each panel the dashed line indicates 
the real age while the symbols and error bars provide the median and inferior and superior uncertainties (1-sigma equivalents) of each distribution.}
\label{agedist1}
\end{figure}

	The Monte Carlo simulations for the $10^4$ \Ms/ clusters with ages between 1 Ma and 10 Ga previously described were used as input for CHORIZOS. Three types of executions depending on the photometry 
and weights used were performed: [a] $UBVRIJHK$ photometry with weights for each filter derived from the age-dependent single-band values of $\theta$ in Table~\ref{theta} (see \citealt{CervLuri07}); 
[b] $UBVRIJHK$ photometry with uniform weights for each filter; [c] $UBV$ photometry with weights derived as in [a]. CHORIZOS was executed for the 10\,000 simulated clusters generated for each real age and 
execution type by comparing the observed colors with the expected photometry of well-populated (i.e. of infinite mass) clusters with ages between 1 Ma and 10 Ga and evaluating the likelihood in each case. 
Such an experiment is a rather realistic simulation of what happens with real data. Note that our solution space is one-dimensional (age) while our data has 2 or 7 free inputs (colors) and that SIMF sampling
effects will force the observed photometry to be incompatible in a strict sense with the available solutions. For each simulated cluster, the observed age was taken to be the mode of the likelihood distribution 
and the combination of the $V$ magnitude and the mode of the observed age was used to derive its observed mass (taking into account the expected evolution of $V$ with age for clusters without SIMF sampling effects). 
The distribution of observed ages (modes) for each real age and execution type is shown in Fig.~\ref{agedist1}. Table~\ref{unkage} gives the median of the observed age and mass distributions, as well as their 
inferior and superior uncertainties derived from their respective distributions. The last column in Table~\ref{unkage} gives the uncertainty in mass derived from the value of $\theta$ for $V$ in the previous 
subsection (i.e. assuming that age and extinction are known).

	I first compare executions [a] and [b]. For MS and LBV clusters, results are nearly undistinguishable. However, for later ages, the weighted executions are significantly better. Unweighted
executions have in general complex distributions in age, with multiple peaks present in most cases. Some of the secondary peaks contain more than 20\% of the clusters and are located at observed ages very
different from the real ones (e.g. log(age) of 6.8 for 1~Ga clusters). The weighted executions with $UBVRIJHK$ photometry, on the other hand have, for the most part, three desirable qualities: they
are [a] nearly-Gaussian with [b] relatively narrow widths, and [c] centered close to the real age. The three (relatively minor) exceptions are 1 Ma clusters (offset\footnote{One point to be considered for 
1 Ma and 10 Ga clusters is that they are located at the edges of the possible age ranges, thus possible offsetting the observed distribution by creating ``one-sided Gaussians''. However, as discussed later on,
this is likely not the source of the offset for 1 Ma clusters.} in median log(age) of 0.25), 31.6 Ma clusters
(non-Gaussian distribution with moderately large width), and 100 Ma (secondary minor peak around log(age) of 6.6). Those three ages will be analyzed in the discussion but let us note here that the unweighted
executions show the same or worse problems at those ages. Therefore, I conclude that {\bf using photometric weights based on the degree of stochasticity of each filter can significantly reduce the
errors in the estimated ages of clusters}.

	I now compare executions [a] and [c]. In general, differences between them are smaller than between [a] and [b] and can be broken down into three age regimes:

\begin{enumerate}
  \item For MS and LBV clusters, differences are minor, as it was also the case between [a] and [b]. The only significant difference is the presence of a few LBV clusters with observed log(ages) 
        between 7.0 and 8.0 for the $UBV$ photometry execution (barely visible at the bottom of the lower plot in Fig.~\ref{agedist1}). 
  \item For RSG and AGB1 clusters, $UBVRIJHK$ photometry has the advantage with respect to the determination of ages. For 10 Ma clusters, the $UBV$ execution shows a strong secondary peak at log(age) of 6.6 
        that is much weaker in the $UBVRIJHK$ execution. For AGB1 clusters, both executions are relatively poor but the $UBVRIJHK$ case has a somewhat wider distribution.
  \item For ages between 100 Ma and 10 Ga, the $UBV$ executions provide better results with narrower widths in the distribution.
\end{enumerate}

	Regarding the measured masses, two results can be derived from Table~\ref{unkage}. First, the median masses for executions [a] and [c] are in most cases similar to each other and a few percent 
lower than the real masses. The most significant differences are for MS clusters (in both executions the median masses deviate from the real ones by $\approx$13\% and the distribution has a long tail toward
even lower masses), RSG clusters (the median mass from [a] is much better than the one from [c]), and AGB1 clusters (the median mass from [c] is much better than the one from [a]). Second, the introduction of
an unknown age does not affect much the dispersion in masses\footnote{In Table~\ref{unkage} I give inferior and superior uncertainties derived from percentiles instead of a single value for the dispersion
(i.e. the standard deviation of the distribution) because some cases show highly asymmetric distributions. Nevertheless, in most cases the standard deviation is approximately the average of the two 
uncertainties given.} for 3.16 Ma and 10 Ma clusters (compared to the dispersion using $V$ photometry for cluster of known age) but for the rest of the ages it increases the dispersion by factors 
between 2 and 3. This single-mass analysis suggests that the value of $\theta$ is left unchanged for LBV and RSG clusters but can increase up to an order of magnitude for clusters of other ages (for a fixed
mass, $\theta$ is proportional to $\sigma^2$). In order to verify such an assumption, we need to analyze the equivalent results to those in Table~\ref{unkage} for a different mass. For that purpose,
I have repeated the CHORIZOS analysis using the Monte Carlo simulations for $10^5$~\Ms/ clusters and verified that increasing the cluster mass by a factor
of ten reduces the dispersion in mass by factors between 3 and 4 (i.e. similar to the expected $\sqrt{10}$), so it appears that, at least in this mass range\footnote{For very low cluster masses, this is
likely to break down because one would expect all distributions in age to become very wide and with multiple peaks.}, the uncertainties in cluster masses with unknown ages approximately satisfy 
Eqn~\ref{sigma2}.

	One preliminary conclusion from this comparison between executions [a] and [c] is that even though adding more filters to the mix has potential benefits, this is not always the case. I will revisit 
this issue when extinction is included. 

\subsection{Unknown ages and extinctions}

\begin{figure}
\centerline{\includegraphics*[width=0.59\linewidth, bb=160 350 445 720]{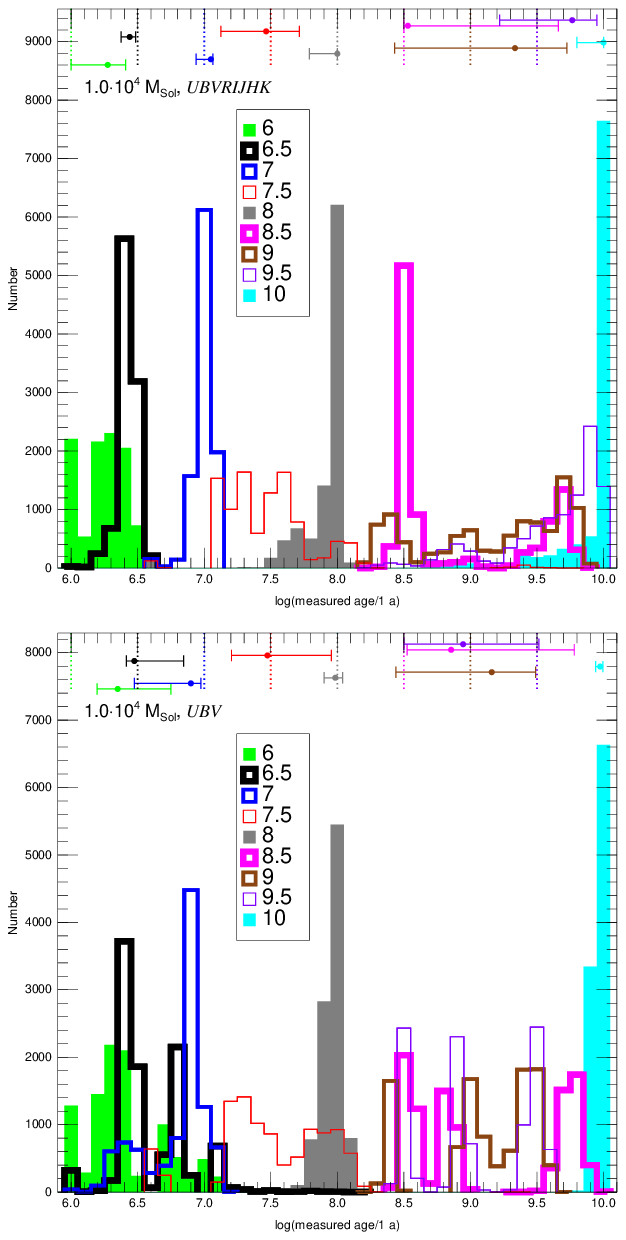}}
\caption{Same as Fig.~\ref{agedist1} for clusters of $10^4$ \Ms/ and unknown extinction but only for the first and last execution types.}
\label{agedist2}
\end{figure}

\begin{figure}
\centerline{\includegraphics*[width=0.59\linewidth, bb=160 350 445 720]{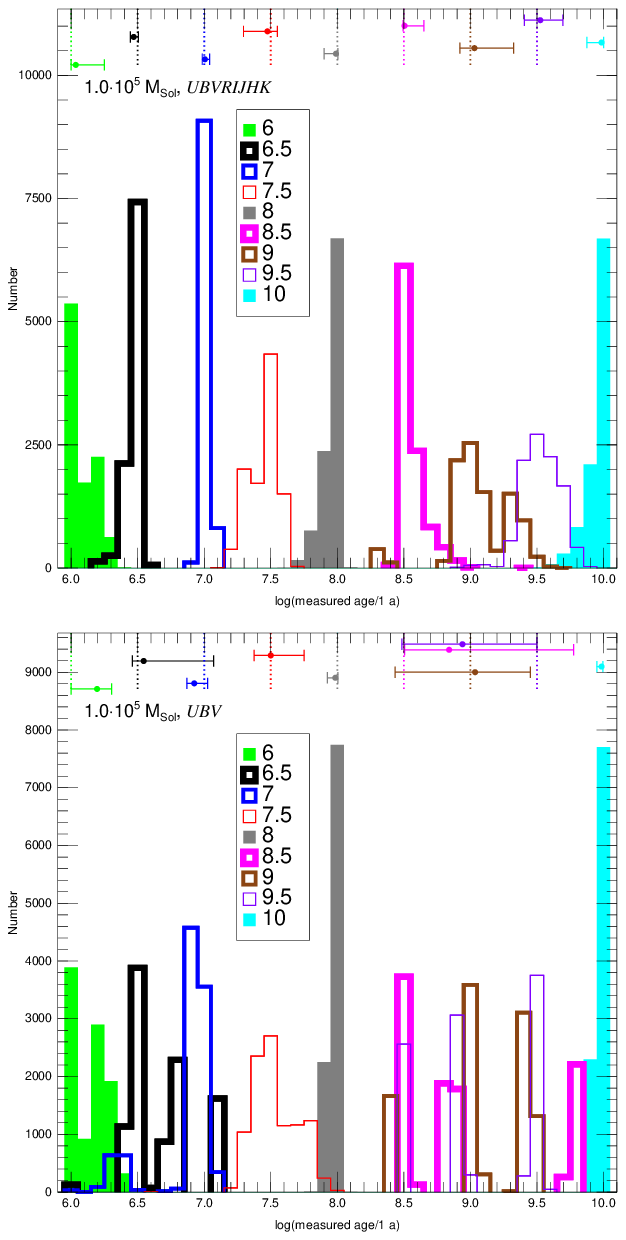}}
\caption{Same as Fig.~\ref{agedist1} for clusters of $10^5$ \Ms/ and unknown extinction but only for the first and last execution types.}
\label{agedist3}
\end{figure}

	As a final step towards simulating a realistic measurement of masses for an ensemble of stellar clusters, I consider the case in which mass, age, and extinction are simultaneously determined from
the photometry. Once again, I use the Monte Carlo simulations for $10^4$~\Ms/ and $10^5$~\Ms/ clusters. Having previously determined that photometric weights based on the degree of stochasticity significantly
reduce the errors, I consider only the two types of CHORIZOS execution in the previous subsection with non-uniform weights (based on $UBVRIJHK$ and $UBV$ photometry, respectively). I extinguished the 
photometry for all the Monte Carlo simulations using a color excess\footnote{\ebv\ is the monochromatic equivalent to $E(B-V)$ and is a direct linear measurement of the amount of dust for a given extinction 
law, as opposed to $E(B-V)$, which is also a function of the input SED and has a non-linear behavior. Similarly, \rv\ is the monochromatic equivalent to $R_V$.} $\ebv = 1.0$ and a \citet{Cardetal89} 
extinction law with $\rv = 3.1$. I executed CHORIZOS in an analogous way to the way it was done in the previous section but this time with two free parameters, log(age) between 6.0 and 10.0 and \ebv\ between 
0.0 and 2.0. The mode of the output 2-D likelihood table as used to select the age and color excess of each cluster. From those two values and the $V$ magnitude the mass was finally calculated. The results
for the ages are shown in graphical form in Figs.~\ref{agedist2}~and~\ref{agedist3} and for the ages, color excesses, and masses for the $10^4$ \Ms/ case in tabular form in Table~\ref{unkageext}.

	As opposed to the case where extinction is known, I find that, for most ages, the $UBVRIJHK$ results are significantly better than those using $UBV$ photometry. $UBV$ photometry
does a particularly bad job of determining the ages for solar metallicity clusters with ages of 31.6 Ma, 316 Ma, 1 Ga, and 3.16 Ga. In those cases multiple peaks are present in the age distribution and 
the observed masses can differ from real ones by large factors. For 100 Ma and 10 Ga clusters $UBV$ photometry does a better job (even slightly better than $UBVRIJHK$) but for the youngest clusters (1 Ma
to 10 Ma) it is significantly worse than $UBVRIJHK$, since the lower plots of Figs.~\ref{agedist2}~and~\ref{agedist3} are the only cases where MS/LBV clusters are misidentified as RSG clusters and
vice-versa. Therefore, I conclude that {\bf the addition of {\it RIJHK} photometry can significantly reduce errors in mass and age if extinction is unknown}.

	The results for clusters between 1 Ma and 100 Ma using $UBVRIJHK$ photometry in Table~\ref{unkageext} are very good: the uncertainties in mass are similar to the reference ones (where both age 
and extinction are known) and only the youngest clusters are slightly biased towards older ages and lighter masses (an effect that we already encountered in the previous subsection). For 100 Ma and 
10 Ga clusters with $10^4$ \Ms/, $UBVRIJHK$ still provides relatively unbiased ages and masses with mass uncertainties only a factor of two larger than the reference ones. The results of lower quality appear
for clusters of 31.6 Ma, 316 Ma, and especially, 1 Ga and 3.16 Ga. For 31.6 Ma clusters of $10^4$ \Ms/, the age distribution is three times wider than the reference one while for 316 Ma clusters of the same
mass a secondary peak is present around 5 Ga. For 1 Ga and 3.16 Ga clusters of $10^4$ \Ms/, the age distribution is very wide and biased; therefore, the age and mass cannot be accurately determined. The
situation is significantly improved for $10^5$ \Ms/ clusters (Fig.~\ref{agedist3}), where mostly unbiased and narrow distributions are observed for all ages using $UBVRIJHK$ photometry (but not so for 
the $UBV$ case). Therefore, if extinction is not independently known, {\it UBVRIJHK} photometry can be used to accurately measure ages and masses of unresolved stellar clusters only for some of the 
interesting regions of the age-mass 2-D space.

\section{Discussion}

	In this paper we have seen that SIMF sampling has large effects in the measurement of stellar cluster masses and the associated CMFs which, if ignored, can lead to substantial biases on the derived 
results. Here I explore under which circumstances it is possible to extract valid and unbiased information about
cluster masses. To that purpose, this discussion is divided into four parts: [a] a description of the cluster mass and age ranges over which masses can be accurately measured, [b] an exploration of some
effects that have not been taken into account in previous sections, [c] a review of the validity of some of the literature results, and [d] guidelines for future work that may extend the measurement ranges
in mass and age.

\subsection{What cluster masses and CMFs can be measured with broad-band photometry?}

	The two most important results of the previous section are the need to give weights to different filters depending on their degree of stochasticity and the advantage of including $RI$ and NIR filters
besides the traditional $UBV$ set when extinction is unknown. Therefore, for this subsection I will assume that $UBVRIJHK$ photometry is used in combination with CHORIZOS or a similar code.

	The most encouraging results are for the youngest clusters, those between 1 Ma and 10 Ma. The values of $\theta$ lie between 500 \Ms/ and 1000 \Ms/ and are relatively impervious to the inclusion of
extinction. Furthermore, even though some mixing is present between 1~Ma and 3.16~Ma clusters for $10^4$ \Ms/ clusters, $UBVRIJHK$ photometry clearly isolates them from older clusters. The mixing between the
two youngest ages takes place (for $10^4$ \Ms/) with a significant fraction of 1~Ma clusters and a much smaller fraction of 3.16~Ma clusters having observed ages around 2~Ma. The reason for the displacement 
of some the youngest clusters towards slightly older observed ages lies in the relative dearth (due to SIMF sampling effects) of very massive stars, which makes those clusters have $U-B$ colors similar to those
of 2 Ma clusters\footnote{Note that there is one technique that is widely used by itself or in combination with broad-band photometry to distinguish ages of young clusters and that is based on the use of
nebular lines to estimate the ionizing flux of the cluster and, from there, its age. See \citet{Maiz00,MacKetal00,Falletal05} for examples.}
(the 2~Ma isochrone is not shown in Fig.~\ref{isoc} but is very similar for most masses to the 1~Ma isochrone, with the only difference taking place near the top of the HRD, where it is slightly
displaced towards lower temperatures). Since a well-populated 2~Ma cluster is brighter than a 1 Ma cluster of the same mass in $V$ (fundamentally due to the strong dependence of the bolometric correction on
temperature), the observed masses for 1~Ma clusters are slightly biased towards lower values in Tables~\ref{unkage}~and~\ref{unkageext}.

	Clusters in the LBV phase show an interesting characteristic: they have high values of $\theta$ for all filters in Table~\ref{theta} but a relatively low value of the same quantity 
when $UBVRIJHK$ photometry is used, even 
when extinction is allowed to vary. The explanation is threefold: [a] the brightest stars in such clusters (mostly B supergiants) have colors which are similar to the integrated cluster colors, so adding or 
subtracting a single bright star changes all of the magnitudes in a similar way; [b] the non-extinguished colors of 3.16 Ma clusters are rather unique (only clusters of 1 Ma have relatively similar colors), 
so it is hard to mistake their ages; and [c] extinction changes their colors in an also unique way. With those properties, the combination of properly-weighted multiple filters can yield a combined value of
$\theta$ lower than the one derived from a single filter such as $V$.

	RSG clusters share some of the properties of LBV clusters. In particular, their colors are unique because they are ``blue in the blue'' and ``red in the red'' i.e. their SEDs are dominated by early-B 
type stars in the optical and by late-type supergiants in the NIR (see \citealt{Maizetal04d} for a practical application of this property). This makes $10^4$~\Ms/ clusters of ages close to 10~Ma relatively 
easy to distinguish from clusters of other ages, even after SIMF sampling effects and extinction are included. In summary, I conclude that {\bf the effects of SIMF sampling on the observed ages and masses of 
young clusters (1~Ma to 10~Ma) determined from broad-band photometry can be relatively minor}. We can use Eqn.~\ref{morange} to quantify this and specify that SIMF sampling effects introduce a change in the CMF 
slope of at most 0.1 for clusters in the MS. LBV, and RSG phases for masses of $10^4$ \Ms/ and higher if $UBVRIJHK$ photometry is used with the appropriate weights.

	The situation changes as we leave the RSG phase. Age uncertainties for AGB1 clusters are several times larger than for 10 Ma clusters of the same age and in most simulations the
age distribution has multiple peaks. The decrease in precision is also noticed in the observed mass, though to a lower degree. Uncertainties decrease as we reach an age of 100 Ma, with values for the age
uncertainties slightly worse than for young clusters and values for the mass uncertainties which are similar or even better. 
For clusters older than 100 Ma, the critical factor is whether extinction is independently known or not. In the
former case, age uncertainties can be kept under control while in the latter they are very large already for $10^4$ \Ms/ clusters. In summary, {\bf the effects of SIMF sampling on the observed ages and masses 
of intermediate-age and old clusters strongly depends on the specific age, filters used, and whether extinction is independently known or not}, so a specific analysis is needed for each observational setup and
circumstances. In some cases it will be possible to keep the uncertainties under control but in others broad-band photometry will be almost useless to determine ages and masses. More specifically, 
{\bf post-RSG clusters (AGB1 phase, log(age) $\sim$ 7.5 Ma) are especially problematic}.

	Even though it is possible to measure ages and masses for clusters in the MS, LBV, and RSG phase, one still has to be careful with contamination from older clusters. That can happen either because 
older ages have broad and smooth observed age distributions that produce an overlap in age or because of the existence of well-defined secondary peaks at the wrong age (e.g. 31.6 and 100 Ma clusters at log(age) of
6.6 and 31.6 Ma clusters at log(age) of 7.1 in the top panel of Fig.~\ref{agedist2}. These secondary peaks can be observed when analyzing real data (see \citealt{Biketal03} for M51), though their exact 
location depends on the used models, because the beginning and end of the RSG phase are a strong function of the metallicity, rotation, and input physics used. The peaks take place at ages where colors change 
rapidly in time, thus creating large unique regions in color space where SIMF sampling effects can move clusters from other ages. The concentration at some ages can also create gaps in others, such as the one
observed around log(age) of 7.1 in \citet{Falletal05}. Those authors indicate that the gap is caused by the introduction of observational errors in the fitting procedure but fail to notice the specific
nature of those errors (SIMF sampling).

\subsection{Other possible problems with the CMF}

	As we have seen, SIMF sampling affects not only the observed masses of individual clusters but also the CMF. I briefly describe here four effects that, when coupled with SIMF sampling, can 
introduce additional biases in the CMF. All of them have to be considered when deriving cluster mass functions from real data.

	The first effect is the magnitude limit. SIMF sampling complicates the derivation of completeness corrections because there is no longer a one-to-one correspondence between luminosity and mass for
a given age. In general, completeness corrections are expected to be smoother functions due to this issue.

	The second effect is the distinction between point sources and extended objects. Some authors (e.g. \citealt{Whitetal99b}) use radial concentration indices to 
separate real clusters from stars. However, such algorithms may identify a cluster dominated by a single star (e.g. a $\sim 10^4$ \Ms/, 3.16 Ma cluster with the ``right'' sampling of the SIMF) as a point
source, hence biasing the derived CMF. Later on I describe one possible way to address this issue.

	The third effect is the possible existence of multiple stellar generations within a cluster or association, a characteristic that has been observed from very young objects (e.g. 30 Dor, \citealt{Walbetal02a})
to old ones \citep{Macketal08} and that invalidates the use of single-burst models (i.e. isochrones) or at least degrades the quality of its results. Such multiple generations are more likely to happen in associations 
or massive compact clusters surrounded by a halo \citep{Maiz01b}. Those objects should be more easily resolved than real, compact clusters, which is one reason for the need of high spatial resolution to 
increase the accuracy of CMF studies. Some studies do not take into account that not all apparent clusters in distant galaxies are real clusters but are instead associations which may contain a significant
fraction of the massive stars \citep{Garm94,ChuGrue08} and that should dissolve rather rapidly, hence providing a simple explanation for the phenomenon of infant mortality \citep{Falletal05}.

	The fourth effect is the evolution of the SMF in a cluster with time due todynamical effects, whih invalidates equating it with the SIMF, even after taking into account stellar mass loss and the existence of 
stellar remnants.  Nowadays it is recognized that there is a continuum in the total (kinetic + potential) energy of stellar groups at birth that spans the zero energy point due to the dominating role of turbulence in 
the star formation process \citep{MacLKles04,Claretal05b}. Furthermore, the cluster may become unbound soon after formation due to gas expulsion \citep{GoodBast06} or later on by intra-cluster stellar encounters or 
external tidal effects \citep{FallRees77}. The loss of stars prior to complete dispersal can modify the observed photometry significantly, especially for old ages \citep{KruiLame08}. The third (multiple stellar
populations in clusters/associations) and the fourth (dynamical evolution) effects are tied together, as evidenced by the difficulty in analyzing the bound character of some stellar groups.

	The fifth effect tied up with SIMF sampling is the validity of the stellar initial mass function itself and of the isochrones used for the comparison between the observed data and the model 
predictions. Since most of the luminosity is produced by the most massive surviving stars in a cluster, the major effect of the discrepancies between the real SIMF and the assumed one (in this paper, Kroupa, see
\citealt{Elme09} for a recent review on SIMF variations) 
should be a normalizing factor for the total mass, which can be analyzed with velocity dispersion studies through the use of the virial theorem (\citealt{KouwdeGr08} and references therein). The validity of the
isochrones is a more complex issue. For the youngest clusters, the two main possible sources of biases are [a] the interplay between rotation, metallicity, and mass loss in the post-MS evolution as a function
of mass \citep{Meynetal08}; and [b] the rejuvenation of the SED induced by mass transfer in close binary systems that allows for the existence of O and/or WR stars beyond the 6 Ma age expected from single-star
evolution \citep{Vanbetal99,MasHCerv99}. However, any of those two issues is unlikely to change the measured ages (either with broad-band photometry or with the equivalent width of Balmer lines) by more than a 
few Ma and should have a relatively small effect in the measured masses. Older clusters pose most significant problems. First, the short-lived post-AGB phase (not included in the isochrones here) is 
sufficient to produce large increases in $\theta$ for the short-wavelength filters because just an individual pAGB star can outshine the rest of the cluster. Second, at low metallicities horizontal-branch stars 
have a similar or larger effect at short wavelengths (and if one goes into the UV, even white dwarfs come into play, see \citealt{Knigetal08} for an example). Third, blue stragglers (the old-age equivalent of 
the rejuvenated O and WR stars previously mentioned) produced by collisions and binary mergers also alter significantly the short wavelength filters and their stochastic behavior. Blue stragglers have the
additional difficulty of strongly depending on dynamical effects, thus complicating the construction of isochrones to model them. In summary, the low values of $\theta$ for $U$ and $B$ in Table~\ref{theta} for
old clusters are not realistic in most circumstances.

\subsection{Testing the validity of literature results}

	Several works have been published in the last decade attempting to measure the CMF and its evolution with time. The results in this paper can be used for a qualitative examination of their validity. 
I discuss\footnote{Only strict SIMF sampling effects plus extinction are dealt with in this subsection; other issues such as distance, model atmospheres, photometric system conversions, metallicity effects,
or analysis code differences are ignored.} three of them here, all of them based on HST/WFPC2 photometry:

\begin{itemize}
  \item \citet{Chanetal99a,Chanetal99b,Chanetal99c} used $UBV$-equivalent + FUV filters to study the stellar clusters in M33 ($d = 0.84$ Mpc).
  \item \citet{Biketal03} (see also \citealt{BoutLame03}) used $UBVRI$-equivalent filters to study the stellar clusters in M51 ($d = 8.4$ Mpc).
  \item \citet{Falletal05} (see also \citealt{Whitetal99b,ZhanFall99}) used $UBVI$-equivalent + H$\alpha$ filters to study the stellar clusters in the Antennae ($d = 22$ Mpc).
\end{itemize}

	A first criticism to all of the three analyses is that none of them uses weights based on the expected behavior of $\theta$. This issue affects especially the M51 and Antennae papers, since they
use a combination of filters that range from $U$-like (F336W) to $I$-like (F814W), which differ greatly in that aspect.

	The observed log(age) for the clusters analyzed by Chandar et al. in M33 range from less than 6.6 to 10.3, similar to the ages used in this paper. Also, most of their analysis is based on 
$UBV$ photometry due to the low detection rate of objects in their sample in the FUV (due to the low sensitivity in that filter and the intrinsic weakness of the FUV luminosity for old clusters). Therefore, 
our third execution type is the relevant comparison. \citet{Chanetal99b} use external extinction determinations based on nearby stars and the overall extinction (both foreground and internal to M33) is low, 
so reddening corrections should be pretty accurate. They find that the majority of their sample has observed masses in the range \sci{4}{2}$-10^4$ \Ms/ and log(age) lower than 8.5. Hence, our results in 
Table~\ref{unkage} and Fig.~\ref{agedist1} are lower limits on the uncertainties induced by SIMF sampling for the M33 clusters. It can be seen that SIMF-related uncertainties are relatively large and should
bias the \citet{Chanetal99b} results. In particular, one would expect that clusters with real log(age) close to 7.5
would suffer especially from SIMF sampling effects and have different observed ages from their real ones, thus biasing the observed age distribution. Indeed, Fig.~10 in \citet{Chanetal99b} shows a minimum close 
to that age and an overall maximum in log(age) between 7.8 and 8.4, which points in that direction.

	Most of the M51 clusters analyzed by \citet{Biketal03} have observed ages below 1 Ga, with less than 1\% of them being apparently older. Their sample is significantly larger than the M33 one and some
of their clusters are more massive but still more than 80\% have masses below $10^4$ \Ms/. They use $UBVRI$ photometry without stochasticity-based weights and simultaneously fit masses, ages, and extinctions. 
For the latter they obtain values of $E(B-V)$ between 0.0 and 1.0. None of execution types in this paper
fits exactly into this description since the ones where extinction is left as a free parameter use either $UBV$ or $UBVRIJHK$ photometry with stochasticity-based weights and the one where uniform weights are 
used is for $UBVRIJHK$ photometry with known extinction. Nevertheless, we can use the following argument: [a] Since the use of uniform weights enhances the uncertainties associated with SIMF sampling, our 
unknown extinction executions (which use stochasticity-based weights) should provide an optimistic limit for the validity of the \citet{Biketal03} results. 
[b] I would expect $UBVRI$ photometry to provide an intermediate case between $UBV$ and $UBVRIJHK$ photometry, so analyzing both executions should give us an insight on the M51 results. With those ideas in
mind, I first find that clusters with real log(age) near 7.5 should once again (as for M33) be scattered in observed log(age) between 6.6 and 8.2. Fig.~13 indeed shows a local minimum between log(age) of 7.5
and 8.0. Also, as noted by the authors, local maxima are found around observed log(age) of 6.70, 7.45, and maybe 7.20. The specific locations of the maxima depend on the code used to generate the synthetic
photometry and, therefore, they should not be overanalyzed. Nevertheless, their existence points towards the existence of SIMF sampling effects in the data. Furthermore, this effect cannot be ignored as being 
of little importance, since, as we have seen, some of the clusters observed at those ages can have real ages which differ by $\sim$1.0 in log(age). The situation for clusters older than 100 Ma is not good 
either. Even though their average mass is larger, Figs.~\ref{agedist2}~ and~\ref{agedist3} show us that when extinction is unknown, it is difficult to measure masses and ages of old clusters from integrated 
broad-band photometry even if their masses are as large as $10^5$ \Ms/.

	The masses of the Antennae clusters in \citet{Falletal05} are significantly higher than the ones in the previous studies, since they analyze clusters of more than \sci{3}{4} \Ms/. They also
incorporate H$\alpha$ observations to their $UBVI$-like photometry which helps in the separation of clusters younger than log(age) lower than 6.6 from older ones. Those characteristics are likely to reduce the
effects of SIMF sampling with respect to the M33 and M51 data. Nevertheless, it is clear from their Fig.~1 (as the authors acknowledge) that the observed age distribution has a nonrandom error component, with
a strong concentration of clusters below log(age) of 7.0 and a dearth between 7.0 and 7.4. As previously discussed, such a distribution is a clear sign of SIMF sampling effects and a look at our
Fig.~\ref{agedist3} (\citealt{Falletal05} fit the extinction from the same data used to calculate ages and masses) show that even for $10^5$ \Ms/ clusters, broad band photometry (with stochasticity-based
weights, the effect should be larger for uniform weights such as the ones used by the authors) still tends to disperse clusters with real log(age) close to 7.5 into a distribution significantly broader than
for clusters with log(age) of 7.0 or 8.0. I suspect that Fig.~2 of \citet{Falletal05} is biased by this effect and that a significant fraction of the clusters in the second age bin in reality belong to the
third.

	Therefore, this paper shows that the existing analyses for M33, M51, and the Antennae likely introduce age biases that produce in turn systematic errors in the age-mass plane. As a consequence, the evolution of 
the CMF with age derived from studies that do not take into account the appropriate biases should be considered with caution. In particular, the region between 10 Ma and 100 Ma is especially subject to SIMF sampling 
effects. This is unfortunate for the sake of evaluating the validity of the Baltimore and Utrecht models, since that is the age range where both models differ more strongly in their predictions (see Fig.~1 in 
\citealt{Lame08}).

\subsection{Possible solutions}

	The reader of the previous subsections may get the impression that using unresolved photometry to obtain accurate ages and masses for stellar cluster systems is a hopeless endeavor. However, I believe 
that is not entirely the case if certain steps are taken. Some of the ideas below are derived directly from the results in this paper while others need further testing:

\begin{enumerate}
  \item {\bf Monte Carlo simulations (such as the ones in this paper) are necessary} and should be performed for the specific filter sets used and ranges in mass and age studied. The results of the
        simulations are needed to estimate the mass and age ranges where results are valid and to generate Bayesian priors (see next point).
  \item {\bf An advanced Bayesian procedure should be applied to the likelihood mass-age map}. Most analyses (including the one in this paper) simply obtain the mode of the likelihood in mass and age (i.e. the
        values that minimize $\chi^2$) and take those values as the observed ones. A further step would be to evaluate the likelihood of all possible values, determine a joint mass-age distribution using a
        reasonable Bayesian prior, derive a combined likelihood for all clusters (i.e. a mass-age distribution function), use the latter as a new Bayesian prior, and iterate as needed. In this way, it may be
        possible to eliminate some of the age biases discussed here, especially in those cases where multimodal solutions exist.
  \item {\bf Weights should be assigned} in such a way that filters with lower stochasticity contribute more to the likelihood map than those with higher stochasticity. Note that in this paper I have used 
        specific weights for each age, which is not possible with real data because we do not know the real ages to start with\footnote{But it is not out of the question to use an iterative procedure fitting
	the observed ages such as the one in the previous point to address this issue.}. However, as it is clear from Table~\ref{theta}, the dependence on wavelength of $\theta$ is rather similar for most ages
	(except for 1 Ma clusters but, as we have seen, age fitting for those clusters is quite robust), so the use of age-averaged weights should not introduce large changes.
  \item {\bf Select the filters before observations are performed}. One advantage of performing the Monte Carlo simulations in advance is that it is possible to select the filters that maximize the information
        on the stellar content. In particular, I hypothesize (but do not prove) that the use of intermediate- and narrow-band filters that measure quantities similar to the Lick indices \citep{Wortetal94} but 
        not necessarily confined to that region of the spectrum (e.g. they could include H$\alpha$, other TiO bands, or the Ca triplet) could lower the values of $\theta$, especially when age and
        extinction need to be disentangled. The use of such filters may be hard to implement because of the simultaneous need for a large collecting area (compared to broad-band photometry) and high spatial 
        resolution over large fields of view and wavelength coverage (which tends to favor space telescopes over ground-based techniques). Intermediate- and narrow-band filters are also helpful in combination
	with the idea in the next point.
  \item {\bf Incorporate the Monte Carlo simulations into the age-fitting code}. This implies carefully selecting a relative large number of SIMF realizations ($\gtrsim$100) for each age and mass range,
        computing the predicted photometry for all of them, and running a CHORIZOS-like code with the realizations as an added dimension. Then, after the likelihood is computed, the additional dimension
	can be collapsed in order to end up with a likelihood map (in e.g. mass, age, and extinction) that can be analyzed with Bayesian techniques. Such an analysis would effectively take into account 
	SIMF sampling effects but is likely to be very costly in computational terms. The use of intermediate- and narrow-band filters would be a nice inclusion into such a code because the measurement of e.g.
	the Ca triplet equivalent width would allow an estimation of the real number of RSGs and AGB stars when the predicted number for their number at a given mass and age is so low that SIMF sampling effects 
	are very large.
  \item {\bf If the clusters can be partially resolved into stars, use the number and colors of the bright point sources to constrain the giant/supergiant population}. See the analysis for NGC 4214 I-Es in
        \citet{Ubedetal07a,Ubedetal07b} for an example. The objections to such an strategy are that it is labor intensive (it requires a one by one detailed effort for each cluster) and that with HST (or 
	similar) resolution it can be extended effectively only up to distances of a few Mpc. Nevertheless, it would be an interesting project to e.g. reprocess the \citet{Chanetal99b} results for M33 with 
	such an strategy and adding information from longer wavelength data obtained with new observations of the same fields.
  \item {\bf Also, for partially resolved clusters and associations, attempt to do separate fits to the existing subclusters.} As previously mentioned, multiple ages associated to different regions in extragalactic 
        stellar clusters have been known to exist for some time now. Therefore, whenever high-spatial-resolution imaging is available, one should try to find age differences between subgroups for young objects.
\end{enumerate}

\section{Conclusions}

\begin{enumerate}
  \item SIMF sampling effects are the largest contributor to the observational uncertainties of the observed masses (and ages) of unresolved clusters studied with broad-band photometry.
  \item A Poisson mass (or equivalent quantity) needs to be calculated for each set of used filters and possible values of the additional parameters (age, extinction\ldots) in order to determine the range
        of validity of the results for the individual cluster masses and the CMF. If multiple filters are used, the most straightforward way to calculate the Poisson mass is with Monte Carlo simulations.
  \item A rule of thumb is that if the real CMF slope is close to $-2.0$ and one desires to calculate the CMF slope with an accuracy of 0.1 without introducing any corrections, then only masses larger than 10 
        times the Poisson mass can be included. 
  \item For most ages, the Poisson mass of a given filter increases with wavelength between $U$ and $K$.
  \item The use of filter-dependent weights based on the variation of the Poisson mass with wavelength can reduce the biases and uncertainties introduced by SIMF sampling.
  \item When using broad-band optical+NIR photometry, it is possible to accurately measure masses and ages of young clusters (MS, LBV, and RSG phases) down to $10^4$ \Ms/ and possibly less.
  \item Post-RSG clusters with ages less than 100 Ma are especially affected by SIMF sampling effects.
  \item The behavior of SIMF sampling effects in old clusters is strongly dependent on whether extinction is known or not and on the possible existence of UV-bright stars (pAGB, HB, blue stragglers).
  \item There are several avenues of improvement that could reduce the uncertainties and biases induced by SIMF sampling.
  \item But one also has to recognize that in the low cluster-mass limit there are cases where it is not possible to unambiguously measure individual cluster masses from unresolved photometry with accuracy 
        because two clusters with different masses can have near-identical luminosities and colors. E.g. two clusters, one with a mass of 100 \Ms/ and another one with a mass of 200 \Ms/, that only differ in the 
        existence in the second one of a 100 \Ms/ star (or of a 60 \Ms/ star and a 40 \Ms/ star) that has already exploded as a SN. In such limit, only an statistical interpretation of the CMF should be possible.
\end{enumerate}

\begin{acknowledgements}

Support for this work was provided by the Spanish Government Ministerio de Ciencia e Innovaci\'on through grants AYA2007-64052 and AYA2007-64712,
the Ram\'on y Cajal Fellowship program, and FEDER funds. I would like to thank Miguel Cervi\~no, Enrique P\'erez, and Rosa Gonz\'alez Delgado for fruitful 
conversations on this topic and an anonymous referee for useful suggestions. I would also like to thank Bernard Plez for granting me access to unpublished MARCS model atmospheres.

\end{acknowledgements}

\bibliographystyle{apj}
\bibliography{biases3}

\appendix
\section{A Bayesian formalism for the comparison between observed and real quantities}

	Suppose that for each member of a population of astronomical objects we want to measure a quantity $X$ (luminosity, mass, distance\ldots) 
and study the properties of its real distribution \fx. An object with a real value of $X$ given by $x$ will be observed to have a value of 
\xo, with $\xo\ne x$ in general. The observed quantity \xo\ may differ from the real one $x$ because of observational or model uncertainties. If
uncertainties can be described by a Gaussian with mean $x$ and standard deviation $\sigma$, then the distribution of \xo\ for a given $x$ (i.e. the conditional 
distribution of the observed quantity) will be given by:

\begin{equation}
\fxocx = \frac{1}{\sqrt{2\pi}\sigma}\exp\left[-\frac{1}{2}\left(\frac{\xo-x}{\sigma}\right)^2\right],
\label{fxocx}
\end{equation}

\noindent the joint distribution of \xo\ and $x$ (i.e. the global distribution in a 2-dimensional observed + real space) by:

\begin{equation}
\fxox = \fxocx \fx = \frac{1}{\sqrt{2\pi}\sigma}\exp\left[-\frac{1}{2}\left(\frac{\xo-x}{\sigma}\right)^2\right] \fx,
\label{fxox}
\end{equation}

\noindent and the observed distribution by:

\begin{equation}
\fxo = \int \fxox \, dx = \int \frac{1}{\sqrt{2\pi}\sigma}\exp\left[-\frac{1}{2}\left(\frac{\xo-x}{\sigma}\right)^2\right] \fx\, dx.
\label{fxo}
\end{equation}

	The combination of [a] \fxo\ being different from \fx\ and [b] the possible existence of selection effects linked to e.g. sensitivity limits is the 
source of a number of biases in astronomy, such as the Lutz-Kelker \citep{LutzKelk73,Smit03} and Malmquist biases \citep{Malm22,Butketal05}. Biases affect
the statistical properties of a sample of observed objects and they depend not only on the observed values and uncertainties (as it is sometimes assumed) 
but also on selection effects and the properties of \fx\ (see \citealt{Maiz01a} and \citealt{Smit03} for the Lutz-Kelker case). If \fx\ is known, it is
possible to compute a correction for individual objects using:

\begin{equation}
c = \bar{x} - \xo = \frac{\int \fxox \,x \, dx}{\int \fxox \, dx} - \xo,
\label{canal3}
\end{equation}

\noindent where I am allowing for the possibility that \fxox\ is not normalized (see \citealt{Maizetal04b,Maizetal08c} for examples of 
applications of such corrections to trigonometric parallaxes). Our best estimate for $X$ would then be $\xo + c$. In practical terms, this is complicated by the 
fact that $c$ itself depends on \fx. One solution is to assume a parameterized functional form for \fx\ with a reasonable initial guess for the 
parameters and to iterate over the full procedure until convergence is reached (see \citealt{Maiz01a,Maiz05c}).

	Besides knowing a correction for the value of an individual object, it is also interesting to know whether the observed distribution at a given
value is larger or smaller than the real distribution i.e. whether uncertainties contribute to increasing or decreasing the number of objects at a certain point
in \xo. This can be evaluated through the fractional number change $p$ between real and observed distributions, given by:

\begin{equation}
p = \frac{\fxo - f_{X}(\xo)}{f_{X}(\xo)} = \frac{\fxo}{f_{X}(\xo)} - 1.
\label{panal3}
\end{equation}

	In this paper I apply the formalism above to the measurement of CMFs from photometric data. I will assume that the real cluster mass 
distribution is given by a power law:

\begin{equation}
\fm = A m^\gamma
\label{fm}
\end{equation}

\noindent between a lower and an upper mass limits \ml\ and \mu, with $A$ being a normalizing factor and $\gamma$ the power law exponent\footnote{Note that 
some papers use $\beta$ for the exponent. Here I prefer $\gamma$ to maintain notation uniformity with the stellar case.}.

	In general, if \fm\ is given by a power law, \fmo\ will be a different function. However, we can still
define a local value for the slope \gammap\ of \fmo\ and a change in slope \dgamma\ between the real and observed distributions as:

\begin{equation}
\dgamma = \gammap - \gamma = \frac{d\fmo/d\mo}{\fmo/\mo} - \gamma.
\label{dgammaanal3}
\end{equation}

\begin{deluxetable}{lcrr}
\tablecaption{Mass uncertainty sources for the Antennae observations described in the text.}
\tablewidth{0pt}
\tablehead{
Type          & Mass dependence     & \multicolumn{2}{c}{$\sigma$ (\Ms/)}               \\
              &                     & $m = 10^3$ \Ms/         & $m = 10^5$ \Ms/
}
\startdata
Photometric   & $m^{0.5}$ (approx.) &     \sci{2.6}{1} $\;\;$ &     \sci{2.2}{2} $\;\;$ \\
Model         & complex             & $<$ \sci{5.0}{1} $\;\;$ & $<$ \sci{5.0}{3} $\;\;$ \\
SIMF sampling & $m^{0.5}$           &     \sci{7.4}{2} $\;\;$ &     \sci{7.4}{3} $\;\;$ \\
\enddata
\label{unctypes}
\end{deluxetable}

\begin{deluxetable}{ccccccrrrr}
\tablecaption{Results of the Monte Carlo simulations of \fmocm\ for 10 Ma clusters observed in $V$ with known age and extinction.}
\tablewidth{0pt}
\tablehead{
$m$        & \ns          & $\overline{m}_{\rm o}$ & $\sigma$       & $\theta$ & $k$             & \sk\;\;\; & \skg\;\; & \ku\;\;\;\; & \kug\;\; \\
(\Ms/)     &              & (\Ms/)                 & (\Ms/)         & (\Ms/)   &                 &           &          &             &
}
\startdata
\sci{3}{2} & \sci{3.3}{5} & \sci{3.01}{2}          & \sci{4.084}{2} & 556.1    & \sci{5.302}{-1} & 3.694     & 2.747    & 17.603      & 11.316   \\
\sci{1}{3} & \sci{1.0}{5} & \sci{1.00}{3}          & \sci{7.494}{2} & 561.6    & \sci{1.767}{+0} & 2.022     & 1.504    &  5.238      &  3.395   \\
\sci{3}{3} & \sci{3.3}{4} & \sci{3.00}{3}          & \sci{1.300}{3} & 563.1    & \sci{5.302}{+0} & 1.186     & 0.869    &  1.785      &  1.132   \\
\sci{1}{4} & \sci{1.0}{4} & \sci{1.00}{4}          & \sci{2.397}{3} & 574.7    & \sci{1.767}{+1} & 0.643     & 0.476    &  0.456      &  0.339   \\
\sci{3}{4} & \sci{1.0}{4} & \sci{3.00}{4}          & \sci{4.133}{3} & 569.4    & \sci{5.302}{+1} & 0.366     & 0.275    &  0.361      &  0.113   \\
\sci{1}{5} & \sci{1.0}{4} & \sci{1.00}{5}          & \sci{7.488}{3} & 560.6    & \sci{1.767}{+2} & 0.202     & 0.150    &  0.038      &  0.034   \\
\sci{3}{5} & \sci{1.0}{4} & \sci{3.00}{5}          & \sci{1.308}{4} & 570.1    & \sci{5.302}{+2} & 0.155     & 0.087    & -0.031      &  0.011   \\
\sci{1}{6} & \sci{1.0}{4} & \sci{1.00}{6}          & \sci{2.389}{4} & 570.9    & \sci{1.767}{+3} & 0.110     & 0.048    &  0.004      &  0.003   \\
\enddata
\label{MonteCarlores}
\end{deluxetable}

\begin{deluxetable}{crrrrrrrrr}
\tablecaption{Values of $\theta$ for different log(ages) (in years) and filters for clusters with known age and extinction.}
\tablewidth{0pt}
\tablehead{
       & \multicolumn{9}{c}{$\theta$ (\Ms/)}                                                     \\
Filter &     6.0 &     6.5 &     7.0 &     7.5 &     8.0 &     8.5 &     9.0 &     9.5 &    10.0
}
\startdata
$U$    &     235 &     945 &     160 &     149 &      53 &      33 &      23 &      26 &      42 \\
$B$    &     200 &  1\,314 &     192 &     305 &      51 &      32 &      27 &      52 &      92 \\
$V$    &     190 &  1\,555 &     566 &     370 &      89 &      38 &      51 &     106 &     163 \\
$R$    &     186 &  1\,722 &  1\,034 &     443 &     151 &      72 &     109 &     173 &     242 \\
$I$    &     178 &  1\,949 &  1\,585 &     616 &     273 &     463 &     476 &     477 &     581 \\
$J$    &     157 &  2\,741 &  2\,240 &  1\,170 &     708 &  3\,598 &  2\,930 &  2\,702 &  2\,445 \\
$H$    &     146 &  3\,081 &  2\,476 &  1\,640 &  1\,281 &  5\,184 &  4\,828 &  4\,499 &  4\,123 \\
$K$    &     140 &  3\,333 &  2\,542 &  1\,771 &  1\,527 &  5\,717 &  5\,688 &  5\,512 &  5\,058 \\
\enddata
\label{theta}
\end{deluxetable}

\begin{deluxetable}{crrrrrrrrr}
\tablecaption{Values of \moLLL\ for different log(ages) (in years) and filters for clusters with known age and extinction.}
\tablewidth{0pt}
\tablehead{
       & \multicolumn{9}{c}{\moLLL (\Ms/)}                                                       \\
Filter &     6.0 &     6.5 &     7.0 &     7.5 &     8.0 &     8.5 &     9.0 &     9.5 &    10.0
}
\startdata
$U$    &  1\,093 &  5\,741 &  1\,745 &     754 &     223 &      96 &      57 &      81 &     175 \\
$B$    &  1\,018 & 13\,477 &  3\,265 &  1\,164 &     288 &     103 &     193 &     315 &     577 \\
$V$    &  1\,000 & 17\,509 &  2\,796 &  1\,192 &     464 &     410 &     551 &     650 &  1\,074 \\
$R$    &     993 & 20\,302 &  2\,795 &  1\,203 &     998 &     896 &     961 &  1\,058 &  1\,524 \\
$I$    &     978 & 23\,911 &  2\,801 &  1\,954 &  2\,755 &  2\,643 &  2\,501 &  2\,825 &  4\,189 \\
$J$    &     930 & 36\,216 &  4\,083 &  3\,894 &  7\,328 &  6\,070 &  8\,796 & 14\,406 & 19\,417 \\
$H$    &     904 & 41\,098 &  4\,570 &  5\,129 & 11\,822 &  7\,386 & 11\,734 & 19\,235 & 26\,623 \\
$K$    &     889 & 44\,401 &  4\,809 &  5\,572 & 13\,891 &  7\,806 & 13\,369 & 21\,949 & 30\,617 \\
\enddata
\label{moLLL}
\end{deluxetable}

\begin{deluxetable}{cr@{\,}lr@{\,}lr@{\,}lrr@{\,}lr@{\,}lr@{\,}lr}
\renewcommand\arraystretch{1.2}
\tablecaption{Ages and masses calculated with CHORIZOS for $10^{4}$ \Ms/ clusters with unknown age and known extinction. The last column gives the expected mass uncertainty for known age and extinction.}
\tabletypesize{\scriptsize}
\tablewidth{0pt}
\tablehead{
Real          & \multicolumn{6}{c}{log(age/1 a)}                                                        & & \multicolumn{6}{c}{$m$ (\Ms/)}                                                          & $\sigma$ (\Ms/) \\
log(age/1 a)  & \multicolumn{2}{c}{7 filt.} & \multicolumn{2}{c}{7 filt.} & \multicolumn{2}{c}{3 filt.} & & \multicolumn{2}{c}{7 filt.} & \multicolumn{2}{c}{7 filt.} & \multicolumn{2}{c}{3 filt.} & $V$             \\
              & &                           & \multicolumn{2}{c}{unweig.} & &                           & & &                           & \multicolumn{2}{c}{unweig.} & &                           &                   
}
\startdata
 6.0 &   $  6.25$&$^{+  0.08}_{-  0.17}$ &   $  6.25$&$^{+  0.08}_{-  0.17}$ &   $  6.25$&$^{+  0.08}_{-  0.19}$ &  &   $  8765$&$^{+  1002}_{-  3204}$ &   $  8753$&$^{+  1013}_{-  3164}$ &   $  8657$&$^{+  1119}_{-  3085}$ & $  1378$ \\
 6.5 &   $  6.45$&$^{+  0.06}_{-  0.05}$ &   $  6.45$&$^{+  0.06}_{-  0.05}$ &   $  6.45$&$^{+  0.06}_{-  0.05}$ &  &   $  9846$&$^{+  3537}_{-  2418}$ &   $  9855$&$^{+  3527}_{-  2420}$ &   $  9773$&$^{+  3631}_{-  2548}$ & $  3943$ \\
 7.0 &   $  7.05$&$^{+  0.04}_{-  0.15}$ &   $  7.05$&$^{+  0.07}_{-  0.04}$ &   $  6.93$&$^{+  0.12}_{-  0.34}$ &  &   $  9558$&$^{+  2003}_{-  1789}$ &   $ 11179$&$^{+  3039}_{-  2573}$ &   $  7879$&$^{+  3311}_{-  5575}$ & $  2379$ \\
 7.5 &   $  7.37$&$^{+  0.32}_{-  0.25}$ &   $  7.29$&$^{+  0.39}_{-  0.18}$ &   $  7.47$&$^{+  0.39}_{-  0.26}$ &  &   $  6514$&$^{+  7009}_{-  2549}$ &   $  5903$&$^{+  7393}_{-  1983}$ &   $  9427$&$^{+  9458}_{-  4960}$ & $  1922$ \\
 8.0 &   $  7.97$&$^{+  0.14}_{-  0.22}$ &   $  7.90$&$^{+  0.19}_{-  1.21}$ &   $  7.99$&$^{+  0.08}_{-  0.14}$ &  &   $  9404$&$^{+  3163}_{-  3166}$ &   $  8558$&$^{+  3712}_{-  7910}$ &   $  9782$&$^{+  2285}_{-  2362}$ & $   945$ \\
 8.5 &   $  8.48$&$^{+  0.07}_{-  0.12}$ &   $  8.48$&$^{+  0.20}_{-  0.48}$ &   $  8.49$&$^{+  0.04}_{-  0.03}$ &  &   $  9604$&$^{+  1676}_{-  1671}$ &   $  9543$&$^{+  4267}_{-  5021}$ &   $  9951$&$^{+  1103}_{-   930}$ & $   620$ \\
 9.0 &   $  8.99$&$^{+  0.09}_{-  0.08}$ &   $  9.00$&$^{+  0.25}_{-  2.18}$ &   $  8.99$&$^{+  0.05}_{-  0.05}$ &  &   $  9789$&$^{+  2537}_{-  1748}$ &   $ 10194$&$^{+  6189}_{- 10027}$ &   $  9852$&$^{+  1962}_{-  1413}$ & $   715$ \\
 9.5 &   $  9.48$&$^{+  0.16}_{-  0.10}$ &   $  9.50$&$^{+  0.25}_{-  0.50}$ &   $  9.49$&$^{+  0.11}_{-  0.08}$ &  &   $  9720$&$^{+  4494}_{-  2784}$ &   $  9756$&$^{+  6754}_{-  6631}$ &   $  9844$&$^{+  3594}_{-  2368}$ & $  1032$ \\
10.0 &   $  9.95$&$^{+  0.05}_{-  0.17}$ &   $  9.94$&$^{+  0.05}_{-  0.44}$ &   $  9.98$&$^{+  0.02}_{-  0.12}$ &  &   $  9003$&$^{+  2172}_{-  3428}$ &   $  8632$&$^{+  2522}_{-  5323}$ &   $  9413$&$^{+  1792}_{-  2869}$ & $  1277$ \\
\enddata
\label{unkage}
\end{deluxetable}

\begin{deluxetable}{cr@{\,}lr@{\,}lrr@{\,}lr@{\,}lrr@{\,}lr@{\,}lr}
\renewcommand\arraystretch{1.2}
\tablecaption{Ages, color excesses, and masses calculated with CHORIZOS for $10^{4}$ \Ms/ clusters with unknown age and extinction. The last column gives the expected mass uncertainty for known age and extinction.}
\tabletypesize{\scriptsize}
\tablewidth{0pt}
\tablehead{
Real          & \multicolumn{4}{c}{log(age/1 a)}                          & & \multicolumn{4}{c}{\ebv}                                  & & \multicolumn{4}{c}{$m$ (\Ms/)}                            & $\sigma$ (\Ms/) \\
log(age/1 a)  & \multicolumn{2}{c}{7 filt.} & \multicolumn{2}{c}{3 filt.} & & \multicolumn{2}{c}{7 filt.} & \multicolumn{2}{c}{3 filt.} & & \multicolumn{2}{c}{7 filt.} & \multicolumn{2}{c}{3 filt.} & $V$             \\
}
\startdata
 6.0 &   $  6.28$&$^{+  0.14}_{-  0.28}$ &   $  6.35$&$^{+  0.40}_{-  0.16}$ &  &         $1.00$&$^{+0.00}_{-0.04}$ &         $0.99$&$^{+0.01}_{-0.54}$ &  &   $  8542$&$^{+  1016}_{-  3466}$ &   $  6806$&$^{+  2894}_{-  2142}$ & $  1378$ \\
 6.5 &   $  6.44$&$^{+  0.05}_{-  0.06}$ &   $  6.47$&$^{+  0.37}_{-  0.06}$ &  &         $1.00$&$^{+0.04}_{-0.01}$ &         $0.99$&$^{+0.02}_{-0.53}$ &  &   $ 10142$&$^{+  4720}_{-  2711}$ &   $ 11031$&$^{+  8596}_{-  3448}$ & $  3943$ \\
 7.0 &   $  7.05$&$^{+  0.01}_{-  0.11}$ &   $  6.90$&$^{+  0.08}_{-  0.43}$ &  &         $0.99$&$^{+0.12}_{-0.15}$ &         $0.98$&$^{+0.32}_{-0.15}$ &  &   $ 10101$&$^{+  2393}_{-  2494}$ &   $  6380$&$^{+  2653}_{-  3374}$ & $  2379$ \\
 7.5 &   $  7.47$&$^{+  0.25}_{-  0.34}$ &   $  7.47$&$^{+  0.48}_{-  0.27}$ &  &         $1.01$&$^{+0.14}_{-0.11}$ &         $1.00$&$^{+0.10}_{-0.10}$ &  &   $  8774$&$^{+  6047}_{-  4335}$ &   $  9319$&$^{+ 11052}_{-  4953}$ & $  1922$ \\
 8.0 &   $  8.00$&$^{+  0.00}_{-  0.21}$ &   $  7.99$&$^{+  0.05}_{-  0.09}$ &  &         $1.00$&$^{+0.08}_{-0.07}$ &         $1.00$&$^{+0.06}_{-0.06}$ &  &   $  9461$&$^{+  1203}_{-  1939}$ &   $  9569$&$^{+   854}_{-   910}$ & $   945$ \\
 8.5 &   $  8.53$&$^{+  1.13}_{-  0.03}$ &   $  8.85$&$^{+  0.93}_{-  0.33}$ &  &         $0.96$&$^{+0.09}_{-0.66}$ &         $0.76$&$^{+0.22}_{-0.50}$ &  &   $ 10807$&$^{+ 83593}_{-  1203}$ &   $ 18376$&$^{+103069}_{-  7893}$ & $   620$ \\
 9.0 &   $  9.34$&$^{+  0.39}_{-  0.90}$ &   $  9.16$&$^{+  0.33}_{-  0.72}$ &  &         $0.80$&$^{+0.59}_{-0.23}$ &         $0.94$&$^{+0.41}_{-0.25}$ &  &   $ 20898$&$^{+ 22283}_{- 17097}$ &   $ 15029$&$^{+ 12787}_{- 11162}$ & $   715$ \\
 9.5 &   $  9.77$&$^{+  0.19}_{-  0.55}$ &   $  8.94$&$^{+  0.57}_{-  0.44}$ &  &         $0.90$&$^{+0.30}_{-0.11}$ &         $1.34$&$^{+0.31}_{-0.35}$ &  &   $ 16520$&$^{+  6497}_{- 10980}$ &   $  3170$&$^{+  6936}_{-  1730}$ & $  1032$ \\
10.0 &   $ 10.00$&$^{+  0.00}_{-  0.20}$ &   $  9.98$&$^{+  0.02}_{-  0.04}$ &  &         $0.97$&$^{+0.14}_{-0.07}$ &         $1.00$&$^{+0.05}_{-0.05}$ &  &   $  9305$&$^{+  1497}_{-  2351}$ &   $  9251$&$^{+  1411}_{-  1241}$ & $  1277$ \\
\enddata
\label{unkageext}
\end{deluxetable}

\end{document}